\documentclass[preprint,showpacs,floats,nofootinbib,superscriptaddress,color,
  showkeys]{revtex4}

\usepackage{bm}
\usepackage{amsmath}
\usepackage{graphicx}
\usepackage{graphics}
\usepackage[dvips]{color}

\newcommand{\be}{\begin{equation}} \newcommand{\ee}{\end{equation}}
\newcommand{\ba}{\begin{eqnarray}} \newcommand{\ea}{\end{eqnarray}}

  \newcommand{\sla}[1]{#1 \!\!\!\slash }
\newcommand{\Tr}{\mbox{\rm Tr}}

\begin{document}

\title{Using Electron Scattering Superscaling to predict Charge-changing
Neutrino Cross Sections in Nuclei}

\author{J.E. Amaro}
\affiliation{Departamento de F\'\i sica Moderna,
Universidad de Granada, 
18071 Granada, SPAIN }
\author{M.B. Barbaro}
\affiliation{Dipartimento di Fisica Teorica,
Universit\`a di Torino and
INFN, Sezione di Torino,
Via P. Giuria 1, 10125 Torino, ITALY }
\author{J.A. Caballero}
\affiliation{Departamento de F\'\i sica At\'omica, Molecular y Nuclear,
Universidad de Sevilla, Apdo. 1065, 41080 Sevilla, SPAIN }
\author{T.W. Donnelly}
\affiliation{Center for Theoretical Physics, Laboratory for Nuclear Science 
and Department of Physics,
Massachusetts Institute of Technology,
Cambridge, MA 02139, USA }
\author{A. Molinari}
\affiliation{Dipartimento di Fisica Teorica,
Universit\`a di Torino and
INFN, Sezione di Torino,
Via P. Giuria 1, 10125 Torino, ITALY }
\author{I. Sick}
\affiliation{Departement f\"ur Physik und Astronomie,
Universit\"at Basel,
CH-4056 Basel, SWITZERLAND } 
\begin{abstract}
  
  Superscaling analyses of few-GeV inclusive electron scattering from nuclei
  are extended to include
  not only quasielastic processes, but now also into the region where 
  $\Delta$-excitation dominates. It is shown that, with reasonable assumptions
  about the basic nuclear scaling function extracted from data and information
 from other studies of the relative roles played by correlation and MEC effects,
  the residual strength in the resonance region can be accounted for through
  an extended scaling analysis. One observes scaling upon assuming that the 
  elementary cross section by which one divides the residual to 
  obtain a new scaling function is dominated by the $N\to\Delta$ transition 
  and employing a new scaling variable which is suited to the resonance region. 
  This yields a good representation of the
  electromagnetic response in both the quasielastic and $\Delta$ regions. The 
  scaling approach is then inverted and predictions are made for charge-changing neutrino 
  reactions  at energies of a few GeV, with focus placed on nuclei which are
  relevant for neutrino oscillation measurements. For this a
  relativistic treatment of the required weak interaction vector and
  axial-vector currents for both quasielastic and $\Delta$-excitation processes
  is presented.
\end{abstract}



\date{\today}

\maketitle

\section{Introduction}
\label{sec:intro}

In recent studies of inclusive electron scattering at intermediate to high
energies from nuclei we have explored various aspects of scaling and
superscaling~\cite{Alberico:1988bv,Day:1990mf,Donnelly:1998xg,Donnelly:1999sw,
  Maieron:2001it,Barbaro:2003ie,Barbaro:1998gu}.  The general procedure used
in such analyses is to divide the experimental inclusive cross section by an
appropriate single-nucleon cross section, having contributions from $Z$
protons and $N$ neutrons with their corresponding electromagnetic form
factors, to obtain a reduced cross section. Here the word ``appropriate''
entails two things. First, the usual analysis in the region of the
quasielastic (QE) peak assumes that the dominant process is elastic scattering
from nucleons in the nuclear ground state followed by quasi-free ejection of
the nucleons from the nucleus, and hence the appropriate single-nucleon form
factors are the elastic ones. Second, the nucleons in the nuclear ground-state
are moving (``Fermi motion'') and accordingly the single-nucleon cross section
used must take this into account. Once one has the reduced cross section it
can be plotted versus one or more appropriately chosen variables: if the
results do not depend on some of these variables and a universal behavior is found one
says that the results scale. Specifically, when the reduced cross section is
plotted versus a well-chosen scaling variable (see below) and no dependence on
the momentum transfer $q$ is observed one says that one has scaling of the
first kind. When no dependence occurs on the momentum scale that characterizes
specific nuclei (essentially the Fermi momentum $k_F$ of a given nuclear
species) one says that one has scaling of the second kind. If both types of
scaling behavior are found one says that superscaling occurs.

At sufficiently high energies we have seen both types of scaling behavior. For
specific nuclei one observes quite good first-kind scaling at excitation
energies below the QE peak, namely, in the so-called scaling region. This is
the familiar $y$-scaling behavior. At energies above the peak, where
nucleon resonances (especially the $\Delta$) are important, this type of
scaling is broken for the total reduced cross section.  On the other hand,
from what data we have where longitudinal-transverse separations have been
made, we know that these scaling violations apparently reside in the transverse
response, but not in the longitudinal. The latter appears to superscale.  In
fact, this is not unexpected, since we know that there are contributions which
do not scale arising from meson-exchange currents
(MEC) plus the correlation effects\footnote{Note that such correlation effects are not all that enter: in particular, even in a factorized approach of the type presented here there are both mean-field and short-range correlations in the initial state which are embodied in a nuclear spectral function and which lead to the scaling function deduced from data, as discussed below.} required by gauge invariance which must be considered together with the MEC~\cite{Amaro02,Amaro:2001xz,Amaro03} and from inelastic scattering from the
nucleons~\cite{Barbaro:2003ie}. As discussed below, it is important to observe that MEC and inelastic contributions are predominantly transverse in the kinematic regions of
interest in the present work.
Note that scaling of the second kind works very well in the scaling region
and, even in the resonance region, is only violated at roughly the 20\% level.

Using these recent studies as a basis we now extend our analysis to encompass both the
QE and $\Delta$ regions. Our approach is the following: taking all of the
high-quality data for two specific nuclei of relevance for our later
discussions, namely for $^{12}$C and $^{16}$O, we proceed as outlined above and begin by performing a comprehensive scaling analysis in the QE region. Using our knowledge of the
experimentally determined longitudinal superscaling function as the starting
point, we work backwards and predict the transverse response one would obtain
strictly from the contributions that are present in that function. In other
words, we reconstruct the part of the transverse cross section that does not
have either MEC effects or inelasticity built in. The net inclusive cross
section so obtained is then, in effect, the QE contribution, except for 
corrections arising from MEC and their associated correlations. 
The next step is to subtract this from the data. What is
left should now be dominated by the inelasticity, and, in particular, when not
too far above the QE peak region, we expect the $\Delta$ resonance to be the
most important contribution.

  From other studies, we expect that a similar procedure can now be followed
for the subtracted results. We again reduce the left-over cross section by
dividing by the appropriate single-nucleon cross section, now for the
$N\to\Delta$ transition, and display the result versus a new scaling variable
in which the kinematics of resonance electro-production are respected. As
discussed later, the results scale quite well, suggesting that this procedure
has indeed identified the dominant contributions not only in the QE region,
but also in the $\Delta$ region. We check our analysis by assembling all of
the pieces obtained via the scaling procedures to produce a total inclusive
cross section which can be compared with data. Overall the answers are very
encouraging, and only for specific kinematics do we see deviations as large as
10--20\%.  We believe that effects from MEC and their associated correlations (which are not incorporated using this
approach) are probably responsible for these residuals.

Having met with success in extending the scaling and superscaling analyses from
the scaling region, through the QE peak region and now into the $\Delta$ region
we are in a position to take a step in a new direction. Since we have the 
scaling functions
and can be sure that, upon being multiplied by the electromagnetic $N\to N$ and
$N\to\Delta$ single-nucleon cross sections, the total nuclear electromagnetic
cross section is quite well reproduced, we can just as well multiply by the
corresponding charge-changing weak interaction $N\to N$ and
$N\to\Delta$ single-nucleon cross sections to obtain predictions for neutrino
reactions in nuclei at similar high-energy kinematics. 

Thus, the second motivation for the present investigation
has been to work backwards to 
predictions for these cross sections with the goal of providing 
high-quality results for use in on-going experimental studies of 
neutrino oscillations at GeV energies~\cite{Oscil,Abl95,Chu97,Amb04,Dra04}. 
These studies are presently being
pursued in the MiniBooNE and K2K/T2K experiments~\cite{Oscil}. Both of these and also the forthcoming MINOS, NOvA and MINERvA
\cite{Abl95,Chu97,Amb04,Dra04} initiatives
involve neutrino energies of several GeV where a fully relativistic
treatment of the neutrino-nucleus scattering is mandatory, but hard to
achieve.  Targets of hydrocarbon or water are involved in the cases of MiniBooNE and 
K2K/T2K, and hence $^{12}$C and $^{16}$O are taken as the focus in the present work.
For the others, iron and lead will also be considered, and in this regard we 
note that, to the degree that scaling of the second kind is reasonably well satisfied, one can focus on the nuclei where the most reliable electron scattering data are available and then subsequently obtain predictions for neutrino reactions not only for those nuclei, but for a wide range of targets. 

Any reliable calculation for neutrino scattering should
first be tested against electron scattering data. Here, instead of using a specific model to describe inclusive electron scattering at relatively high energies in the QE and $\Delta$ regions, as stated above, we follow a different approach. Based on the scaling behavior of the
electron-nucleus cross section in both the QE and $\Delta$ peaks we extract
the scaling functions directly from experiment and use them to predict the
neutrino-nucleus cross section. 

This strategy is motivated by the fact that, while relativistic 
modeling~\cite{Horowitz:rj,Kim95,Barbaro96,Amaro96,Kim1996,Alberico97plus,Alvarez-Ruso:1998hi,Maieron:2003df,Bleve:2000hc,Meucci:2003cv} of the 
nuclear dynamics in studies of high-energy inclusive lepton scattering is expected
to be capable of getting the basic size and shape of the cross section, so far it
has not been capable of accounting for important details of the response. Specifically,
such modeling has been able to provide a reasonable representation of the $eA$ 
inclusive cross section. Typically at high energies the cross sections obtained 
using wide classes of models, including those with relativistic mean-field dynamics 
and RPA-type correlations included, are seen to be very similar to the results found 
using the Relativistic Fermi Gas (RFG) model and accordingly we will also make
comparisons of results obtained using the scaling approach with those obtained
using the RFG model. At the peaks of the QE or $\Delta$ responses one finds that
the two approaches differ by about 25\% (with mean-field effects this discrepancy
is reduced to perhaps 20\%); however, as we shall see later, the phenomenological
scaling approach requires a long tail which is largely absent in most modeling. 
A possible reason for this 
disagreement is the absence of classes of short-range correlation effects in 
most of the relativistic modeling.  It goes without saying that non-relativistic 
modeling is completely inadequate at the energies of interest in the present work. 
At intermediate energies (below those considered in this study) it is, of course, 
important to include effects from final-state interactions and RPA 
correlations (see for instance recent work reported in~\cite{QQ1,QQ2}) as these can be significant.

Since most semi-leptonic scattering processes at similar kinematics have much the same character, one should expect that failure at this level will also imply a similar level of disagreement for predictions made of neutrino reaction cross sections using the same types of modeling. Clearly, on the one hand, if all one wants is a rough estimate of neutrino reaction cross sections at similar energies, then existing relativistic modeling is probably adequate. On the other hand, if uncertainties of less than 25\% are required (as, for example, when one wishes to see distortions in the energy distributions of the detected muons in $(\nu_\mu,\mu^-)$ reactions with nuclei caused by neutrino oscillations), then one must use existing models with great caution. As we shall see below, it appears that the superscaling approach being followed in the present work is capable of reducing the uncertainty to perhaps the 10\% level, at least when one  limits the focus to the QE region and the region up to inelasticities where the $\Delta$ contribution reaches its maximum.

The paper is organized the following way. In Sec.~\ref{sec:kine} we begin with
a brief discussion of kinematics, since we will be interested not only in
electron neutrino induced reactions, where the lepton masses can safely be
ignored, but also in muon neutrino induced reactions where the energies,
although relatively high, are not high enough to safely ignore the muon mass.
In Sec.~\ref{sec:supers} we present a summary of the formalism needed in
studies of scaling and superscaling, both for the QE and $\Delta$ regions.
There we give the results of our analysis of inclusive high-energy inelastic 
electron scattering data for carbon and oxygen, using the
procedures outlined above, and thereby validate the scaling functions we use
in the rest of the paper. In Sec.~\ref{sec:ccnureac} we turn to the second
theme of the paper and discuss charge-changing neutrino reactions with nuclei.
We begin by presenting the basic formalism required in treating electroweak
processes, followed by development of the single-nucleon responses in the QE
region (Sec.~\ref{sec:snqe}) and in the $\Delta$ region
(Sec.~\ref{sec:sndel}), now of course with both vector and axial-vector $N\to
N$ and $N\to\Delta$ currents. Section~\ref{sec:cross} then contains a
discussion of the formalism involved in obtaining the cross sections and
response functions. Once these developments are in hand, we proceed in
Sec.~\ref{sec:results} with a presentation of our predictions for
charge-changing neutrino reactions with nuclei. Finally, in
Sec.~\ref{sec:concl} we summarize our work and present our conclusions.

\section{Lepton Scattering Kinematics}
\label{sec:kine}

In this section we begin with a brief discussion of the kinematics involved in
studies of lepton scattering from nuclei, including electron scattering and
the subject of Sec.~\ref{sec:ccnureac}, charge-changing neutrino reactions.
We start with a general scattering problem in which an incident beam of
leptons with 4-momentum $K^{\mu }=(\epsilon ,\mathbf{k})$ scatters and a
lepton with 4-momentum $K^{\prime \mu }=(\epsilon ^{\prime
},\mathbf{k}^{\prime })$ emerges.  In general one has
\begin{eqnarray}
\epsilon &=&\sqrt{m^{2}+k^{2}}  \label{xx1a} \\
\epsilon ^{\prime } &=&\sqrt{m^{\prime 2}+k^{\prime 2}},  \label{xx1b}
\end{eqnarray}
where $m$ and $m^{\prime }$ are the masses of the incident and outgoing
leptons, respectively. Clearly, for electron scattering $m=m^{\prime }=m_{e}$
(usually, but not always, this can be taken to be zero) and for electron
neutrino induced charge-changing neutrino reactions $m=m_{\nu e}\cong 0$,
whereas $m^{\prime }=m_{e}$ (again, essentially zero). The difficult case is
for muon neutrino induced charge-changing neutrino reactions where $m=m_{\nu
\mu }\cong 0$, whereas $m^{\prime }=m_{\mu }$; the last is clearly not
negligible for the kinematics of interest in the present work.

As usual one has a 4-momentum transfer $Q^{\mu }=(\omega ,\mathbf{q})$ with 
\begin{eqnarray}
\omega &=&\epsilon -\epsilon ^{\prime }  \label{xx2a} \\
\mathbf{q} &=&\mathbf{k}-\mathbf{k}^{\prime },  \label{xx2b}
\end{eqnarray}
the energy transfer and 3-momentum transfer, respectively.
The momentum transfer is spacelike: $-Q^{2}=q^{2}-\omega ^{2}>0$. For
convenience we define an average leptonic mass 
\begin{equation}
M\equiv \sqrt{\frac{1}{2}\left( m^{2}+m^{\prime 2}\right) }\geq 0
\label{xx3}
\end{equation}
and, given an excitation from target rest mass $M_{i}$ to some final rest
mass $M_{f}\geq M_{i}$ (that is, the final hadronic rest frame total energy
is $W=M_{f}$), define a sort of excitation energy 
\begin{equation}
\omega _{0}\equiv \frac{1}{2M_{i}}\left( M_{f}^{2}-M_{i}^{2}\right) \geq 0.
\label{xx4}
\end{equation}
Then from energy-momentum conservation one has 
\begin{equation}
\omega =\omega _{0}+\frac{|Q^{2}|}{2M_{i}}.  \label{xx5}
\end{equation}
Solving Eqs.~(\ref{xx2a}) and (\ref{xx5}) together one can get expressions for
the scattered lepton's energy and 3-momentum. Defining 
\begin{eqnarray}
\epsilon _{1} &\equiv &\sqrt{M_{i}^{2}+2M_{i}\epsilon +m^{2}+k^{2}\sin
^{2}\theta }  \label{xx6a} \\
\epsilon _{2} &\equiv &\sqrt{M_{i}(\epsilon -\omega _{0})+M^{2}},
\label{xx6b}
\end{eqnarray}
where $\theta $ is lepton scattering angle (the angle between $\mathbf{k}$
and $\mathbf{k}^{\prime }$), it can be shown that 
\begin{eqnarray}
k^{\prime } &=&\frac{1}{\epsilon _{1}^{2}}\left[ \epsilon _{2}^{2}(k\cos
\theta )+(M_{i}+\epsilon )\sqrt{\epsilon _{2}^{4}-m^{\prime 2}\epsilon
_{1}^{2}}\right]  \label{xx7a} \\
\epsilon ^{\prime } &=&\frac{1}{\epsilon _{1}^{2}}\left[ \epsilon
_{2}^{2}(M_{i}+\epsilon )+(k\cos \theta )\sqrt{\epsilon _{2}^{4}-m^{\prime
2}\epsilon _{1}^{2}}\right] ,  \label{xx7b}
\end{eqnarray}
where for the results to be real for all scattering angles the beam energy
must be greater than $\epsilon _{min}$, where 
\begin{equation}
\epsilon _{min}=m^{\prime }+\omega _{0}+\frac{m^{\prime }\omega
_{0}+(m^{\prime 2}-m^{2})/2}{M_{i}-m^{\prime }}.  \label{xx8}
\end{equation}
Hence, for a given excitation energy $\omega_0$ and for given beam energy
and scattering angle, the quantities $\epsilon_{1,2}$ can be computed and
through them the final lepton's energy and 3-momentum are fixed. Clearly
the 4-momentum transfer is then given as well.

\section{Scaling and Superscaling}
\label{sec:supers}

\subsection{Scaling in the quasielastic peak region}
\label{sec:scalqe}

In this subsection we briefly review the structure of the nuclear 
responses in the region of the quasielastic peak (QEP). We begin with
the basic Relativistic Fermi Gas model that has been used to 
motivate scaling and superscaling behavior in this region of 
kinematics~\cite{Alberico:1988bv,Donnelly:1998xg,Donnelly:1999sw,Day:1990mf}. 
Here a single parameter characterizes the
dynamics, namely, the Fermi momentum $k_F$. In the present work our
goal is to use the electron scattering cross sections as input, perform
a scaling analysis, and arrive at predictions for the charge-changing
neutrino cross sections. Accordingly the focus is placed on kinematic
regimes where the cross sections (induced by electrons or neutrinos) 
are substantial, and from past work it is known that under such 
circumstances it is 
a good approximation to work only to leading order in an expansion in
$\eta_F \equiv k_F/m_N$. Also 
$\xi_F\equiv \sqrt{1+\eta_F^2}-1\cong \eta _{F}^{2}/2$ is small.

The leading-order QE responses (denoted by subscript zero) 
may be written, in the non-Pauli-blocked domain, in the following form~\cite{Alberico:1988bv}: 
\begin{eqnarray}
R_{L}^{QE}(\kappa ,\lambda )_{0} &=&\Lambda_0\frac{\kappa ^{2}}{\tau }\left[ \left(
1+\tau \right) W_{2}(\tau )-W_{1}(\tau )\right] \times f_{RFG}(\psi )
\label{x22a} \\
R_{T}^{QE}(\kappa ,\lambda )_{0} &=&\Lambda_0\left[ 2W_{1}(\tau )\right] \times
f_{RFG}(\psi ),  \label{x22b}
\end{eqnarray}
with
\begin{eqnarray}
\Lambda_0 &\equiv &\frac{{\cal N} \xi _{F}}{m_{N}\kappa \eta _{F}^{3}}
\cong \frac{\cal N}{2\kappa k_{F}},  \label{x20b}
\end{eqnarray}
and $W_1, W_2$ the structure functions for elastic scattering. 
As usual the proton (${\cal N}=Z$) and neutron (${\cal N}=N$) contributions should be separately 
computed with the appropriate form factors  and added together.
The latter are linked to the Sachs form factors through the 
well-known relations
\begin{eqnarray}
\left( 1+\tau \right) W_{2}(\tau )-W_{1}(\tau ) &=&G_{E}^{2}(\tau )
\label{y23a} \\
2W_{1}(\tau ) &=&2\tau G_{M}^{2}(\tau )~.  \label{y23b}
\end{eqnarray}
As usual, we employ dimensionless variables 
$\lambda \equiv\omega /2m_{N}$, $\kappa \equiv q/2m_{N}$ and $\tau
\equiv|Q^{2}|/4m_{N}^{2}=\kappa ^{2}-\lambda ^{2}$. 
The RFG has the following universal form for the superscaling function:
\begin{equation}
f_{RFG}(\psi )=\frac{3}{4}(1-\psi ^{2})\theta (1-\psi ^{2}),  \label{x21}
\end{equation}
that is, when plotted against the scaling variable $\psi$,
\begin{equation}
\psi\equiv \frac{1}{\sqrt{\xi_F}} \frac{\lambda-\tau}{\sqrt{(1+\lambda)\tau+
\kappa\sqrt{\tau(1+\tau)}}},
\label{eq:psi}
\end{equation} 
a universal
behavior is obtained with no dependence left either on momentum transfer
(scaling of the first kind) or on nuclear species via $k_F$ (scaling of the
second kind). 

In studies of electron scattering scaling one usually
includes a small energy shift by replacing $\omega$ by $\omega-E_{shift}$
in order to force the maximum of the QE response to occur for 
$\psi^\prime=0$ (see, for example, 
\cite{Cenni:1996zh,Donnelly:1998xg,Donnelly:1999sw,Maieron:2001it}). 
This is equivalent to taking $\lambda\to\lambda'=
\lambda-\lambda_{shift}$ with $\lambda_{shift}=E_{shift}/2m_N$ and
correspondingly $\tau\to\tau'=\kappa^2-\lambda'^2$ in Eq.~(\ref{eq:psi}). 
Often we shall use the pair of variables $(\kappa,\psi')$ in place of
the original pair $(q,\omega)$ to characterize the inclusive scattering responses ---
clearly they are functionally related by the above equations.

Using the guidance provided by the RFG, the procedure to be adopted in order
to get the experimental scaling function $F^{QE}(\kappa ,\psi')$ in the QE
domain is then clear: one simply divides the experimental QE cross section by
\begin{equation}
S^{QE}\equiv \sigma _{M}\left[ v_{L}G_{L}^{QE}+v_{T}G_{T}^{QE}\right] ,
\label{x24}
\end{equation}
where $\sigma_M$ is the Mott cross section, $v_L$, $v_T$ are the kinematic
factors defined below in Eqs.~(\ref{xx22d},\ref{xx22e}) and 
where the functions $G_{L,T}^{QE}$ 
are~\cite{Day:1990mf,Donnelly:1998xg,Donnelly:1999sw,Maieron:2001it,
Barbaro:2003ie}
\begin{eqnarray}
G_{L}^{QE} &=&\frac{\kappa }{2\tau }\left[ ZG_{Ep}^{2}+NG_{En}^{2}\right] +
{\cal O}(\eta _{F}^{2})  \label{x25a} \\
G_{T}^{QE} &=&\frac{\tau }{\kappa }\left[ ZG_{Mp}^{2}+NG_{Mn}^{2}\right] +
{\cal O}(\eta _{F}^{2}).  \label{x25b}
\end{eqnarray}
The factors involving $\kappa $ and $\tau $ in Eqs.~(\ref{x25a}) and
(\ref{x25b}) arise partly from the Jacobian of the transformation from
$\lambda$ to $\psi$~\cite{Barbaro:1998gu} and partly from the explicit 
calculation leading to Eqs.~(\ref{x22a}) and (\ref{x22b}).
Finally, as in past discussions of scaling of the second kind,
one multiplies $F^{QE}(\kappa,\psi' )$ by $k_{F}$
to obtain the superscaling function $f^{QE}(\kappa ,\psi' )$.

The nuclear response functions all have the general structure
\begin{eqnarray}
\left[ R\right] ^{QE} &=&\frac{1}{k_{F}}f^{QE}(\kappa,\psi' )\frac{%
{\cal N}}{2\kappa } \left[ R \right]^{s.n.} 
\label{xy34a}
\end{eqnarray}
where ${\cal N}$ is the appropriate nucleon number. In particular, as
stated above, one copy of this expression with proton form factors and 
${\cal N}=Z$ should be added to another with neutron form factors and
${\cal N}=N$ for electron scattering. Here $\left[ R \right]^{s.n.}$ is the 
corresponding single-nucleon response.

In previous work \cite{Day:1990mf,Donnelly:1998xg,Donnelly:1999sw} 
we have shown that scaling works quite  well for all nuclei and for
energy loss
$\omega < \omega_{QEP}$ where $\omega_{QEP}$ corresponds to the maximum 
of the quasielastic peak --- the scaling function $F^{QE}(\kappa ,\psi')$
is indeed largely independent of the momentum transfer as long as $q$ is of the
order $2k_F$ or larger. Deviations from scaling, which mainly occur at larger 
energy loss, are related to contributions beyond quasielastic scattering such
as those from meson exchange currents and $\Delta$-excitation (see below).

The scaling behavior  becomes particularly clear if one studies the
experimental response separated into its longitudinal (charge) and transverse
(magnetic) pieces. The non-scaling contributions mentioned above mainly occur
in the transverse response. Accordingly, one finds that the experimental
longitudinal responses scale much better, and to much larger energy loss. The
approach taken in \cite{Donnelly:1998xg,Donnelly:1999sw} therefore has been to
use the experimental longitudinal responses to define the scaling function
$f^{QE}(\psi')$. 

The total inclusive electron scattering response is then assumed to be composed of several contributions: (1) there is the entire longitudinal contribution which appears to superscale and to be represented by the empirical scaling function $f^{QE}(\psi')$; (2) part of the transverse response arises from quasielastic knockout of nucleons from the nucleus and is also driven by the scaling function $f^{QE}(\psi')$; however, (3) the transverse response has additional contributions, at least from MEC effects with their associated correlations and from inelastic single-nucleon processes including the excitation of the $\Delta$. From our past work we know that typically the effects under item (3) break the scaling. The contributions from MEC together with their attendant correlations enter roughly 
at the 10\% 
level~\cite{Amaro02,Amaro:2001xz,Amaro03,Amaro98,DePace03,DePace04,Carlson02,Ama03,Ama94} and, 
as we argue later, may be less important for charge-changing neutrino reactions than they are for electron scattering --- accordingly we shall ignore these non-scaling effects in the present work. In future work we hope to address this issue more directly with continued relativistic modeling of these contributions. Other effects can enter into the dynamics and invalidate the picture here (for example, at low $q$ RPA correlation effects can modify the longitudinal and transverse responses in different ways because of the very different isospin character of these two channels); however, again, at the kinematics of interest in the present work these effects are thought to be relatively small.

The largest non-scaling contribution to the transverse response is then believed to be the one arising from inelastic, but impulsive processes, especially via the excitation of the $\Delta$ for the kinematics of interest in the current study, and this provides the focus for the following subsection.

In~\cite{Donnelly:1998xg,Donnelly:1999sw} the 
inter-comparison of the scaling functions for various nuclei has
been performed in terms of the functions $f^{QE}(\kappa,\psi')$ extracted from 
$F^{QE}(\kappa,\psi')$ as discussed above. Excellent scaling of
the second kind, {\it i.e.,} scaling functions $f^{QE}(\kappa,\psi')$ that 
closely match for different nuclei, was observed and, indeed, such second-kind scaling is actually significantly better realized than is scaling of the first kind.
  
The combination of the scaling of the first and second kind --- superscaling
--- allows one to determine from the data a universal scaling function
$f^{QE}(\psi')$.  The scaling function (and quasielastic cross section) for
individual nuclei can then be recalculated once the Fermi momentum of the
nucleus is known.

\begin{figure}[hbt]
\begin{center}
\includegraphics[scale=0.5,clip,angle=0]{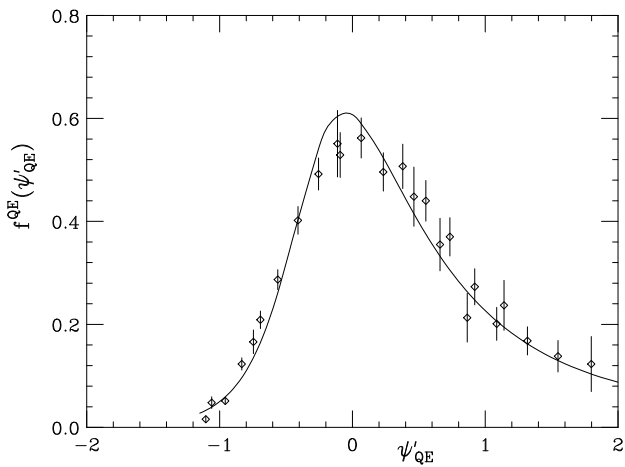}
\parbox{13cm}{\caption[]{
 Averaged experimental $f^{QE}(\psi^\prime_{QE})$ versus 
$\psi^\prime_{QE}$ in the quasielastic region together with a
 phenomenological parameterization of the data. The integral
of the curve has been normalized to unity.
}\label{fpsiplaf}} 
\end{center} 
\end{figure}
Reliable separations of data into their longitudinal and transverse
contributions for $A>4$ are available only for a few
nuclei~\cite{Jourdan:1996}; all of these response functions have been used to
extract the ``universal'' quasielastic response function $f^{QE}$ and
to obtain a parameterization by a simple function. Figure~\ref{fpsiplaf} shows
$f^{QE}(\psi'_{QE})$ averaged over the nuclei employed, together with the
corresponding fit\footnote{In the figure and henceforth the QE
scaling variable is denoted $\psi^\prime_{QE}$ to distinguish it 
from the scaling variable used in the $\Delta$ region; see below.}.  Note 
that $f^{QE}(\psi'_{QE})$ 
has a somewhat asymmetric shape and a tail that extends towards positive values of
$\psi'_{QE}$. In contrast, the RFG (see Eq.~(\ref{x21})) is symmetric when plotted as in the figure, is limited strictly to the region $-1\leq \psi^\prime_{QE}\leq +1$ and has a maximum value of 3/4, while the empirical scaling function reaches only about 0.6. 

One source for the difference could be typical mean-field dynamics in the initial and final nuclear states involved; however, for the kinematics of interest in the present work, both relativistic mean-field theory~\cite{Kim95,Kim1996,Alberico97plus,Maieron:2003df} and relativized shell-model
studies~\cite{Amaro96} appear to provide only rather modest differences from the RFG predictions. As stated above, non-relativistic modeling is quite incapable for such kinematics (see, for instance, \cite{Amaro96}). 

Another source for the differences seen between the RFG or mean-field descriptions and the empirically determined scaling function arises from high-momentum components in realistic wave functions which may be large enough to produce the results shown in the figure, although much more work, especially in a relativistic context, is required to put this on solid footing. For the present we limit our approach to phenomenology and take the scaling function from experiment. It should be emphasized at this point that much of the inability of typical modeling to account for the inclusive response at these kinematics appears to stem directly from the inability of that modeling to account for the results in the figure. We shall see below that when the empirical scaling function is used one obtains a good representation of measured $(e,e')$ cross sections and, therefore, that one's confidence in proceeding to predictions for neutrino reaction cross sections must be raised.

Unseparated experimental quasielastic cross sections 
have been measured for several nuclei over a large range of momentum
transfer (for a compilation see \cite{Donnelly:1999sw}). These data have been
used to determine the values of $k_F$ for the nuclei considered
(in~\cite{Maieron:2001it} a table with numerical values is given) and, since
the evolution of $k_F$ with the nuclear mass number is slow, these values can
easily be interpolated for any nucleus of interest.

As was pointed out above, the description of the experimental scaling functions
involves, besides the choice of the proper $k_F$ that sets the overall momentum
scale, the use of an energy shift $E_{shift}$ that in an average way accounts 
for the nucleon removal energy. This small correction  has been included in 
\cite{Donnelly:1998xg,Donnelly:1999sw} 
in the analysis of the data, and \cite{Maieron:2001it} gives a table 
with the numerical values for various nuclei.     

We should add that, in order to obtain $f^{QE}$ from the cross sections, one
had to divide by the single-nucleon cross sections obtained from $e$-$p$ and
$e$-$n$ scattering. We have used the H\"{o}hler parameterization 8.2
\cite{Hoehler:1976}.  In the range of $q$ of interest here, this
parameterization agrees with more recent parameterizations fitted to a
somewhat more extensive data set.
%>>>>>>>>>>>>>>>>>>>>>>>>>

\subsection{Scaling in the region of the $\Delta$ peak}
\label{sec:scaldel}

Following the framework
of~\cite{Amo97,Amaro:1999be,Maieron:2001it}, let $m_{*}$ be the
mass of a generic nucleon excitation and $\mu _{*}\equiv m_{*}/m_{N}$; hence
$\mu _{*}=1$ for quasielastic scattering and $\mu _{*}=m_{\Delta }/m_{N}\equiv
\mu _{\Delta }$ for electro-excitation in the $\Delta $ region. Introducing
\begin{eqnarray}
\beta _{*} &\equiv &\frac{1}{4}\left( \mu _{*}^{2}-1\right)  \label{x2a} \\
\rho _{*} &\equiv &1+\beta _{*}/\tau  \label{x2b}
\end{eqnarray}
we generalize the dimensionless scaling variable of the quasielastic 
peak as follows:
\begin{equation}
\psi _{*}\equiv \left[ \frac{1}{\xi _{F}}\left( \kappa \sqrt{\rho
_{*}^{2}+1/\tau }-\lambda \rho _{*}-1\right) \right] ^{1/2}\times \left\{ 
\begin{array}{cc}
+1 & \lambda \geq \lambda _{*}^{0} \\ 
-1 & \lambda \leq \lambda _{*}^{0}
\end{array}
\right. ,  \label{x3}
\end{equation}
%where 
%\begin{eqnarray}
%\xi _{F} &\equiv &\sqrt{1+\eta _{F}^{2}}-1 ~,  \label{x4b}
%\end{eqnarray}
which vanishes for 
\begin{equation}
\lambda=
\lambda _{*}^{0}=\frac{1}{2}\left[ \sqrt{\mu _{*}^{2}+4\kappa ^{2}}-1\right]
,  \label{x5}
\end{equation}
or, in dimensionful variables, when 
\begin{equation}
\omega=\omega _{*}^{0}=\sqrt{m_{*}^{2}+q^{2}}-m_{N}~.  \label{x6}
\end{equation}
When $\mu _{*}=1$ one recovers the QE answer in Eq.~(\ref{eq:psi}), where $\beta _{*}=0$, 
$\rho_{*}=1$ and at the peak $\omega_{QE}^{0}=\sqrt{m_{N}^{2}+q^{2}}-m_{N}$. 
As in the previous subsection where the QE scaling variable was discussed, here also we include the small energy shift $E_{shift}$ by making the replacement
$\omega\to\omega'\equiv \omega - E_{shift}$ with $\lambda\to\lambda'$ and 
$\tau\to\tau'$ as before. Again these replacements are made in the above equations to yield a generic shifted scaling variable $\psi'_*$, and specifically for use in the $\Delta$ region the shifted scaling variable
$\psi'_\Delta$.

When considering the $N\to\Delta$ transition structure functions we change
notation from the general quantities $\beta_*$, $\rho_*$, $\psi_*$, 
{\it  etc.,} to $\beta_\Delta$, $\rho_\Delta$, $\psi_\Delta$, {\it etc.,} 
and in addition introduce
\begin{equation}
\gamma_{\Delta }\equiv \frac{1}{4}\left( \mu _{\Delta }-1\right) ^{2},
\label{x10}
\end{equation}
and 
\begin{equation}
\kappa_{\Delta }^{*}=\frac{1}{\mu _{\Delta }}\sqrt{\tau +\left( \tau +\beta
_{\Delta }\right) ^{2}}~,  \label{x12}
\end{equation}
which allow us to define
\begin{eqnarray}
\nu _{1}^{\Delta } &\equiv &
\left( 1+\mu _{\Delta }\right) ^{2}\left( \tau +\gamma _{\Delta }\right)
\label{x13b}
\end{eqnarray}
and 
\begin{equation}
\nu _{2}^{\Delta }\equiv \nu _{1}^{\Delta }\frac{\tau }{\left( \mu _{\Delta
}\kappa _{\Delta }^{*}\right) ^{2}}~.  \label{x14}
\end{equation}
Then the $N\rightarrow \Delta $ single-baryon responses will read:
\begin{eqnarray}
w_{1}^{\Delta }(\tau ) &=&\nu _{1}^{\Delta }\left[ G_{M,\Delta }^{2}(\tau
)+3G_{E,\Delta }^{2}(\tau )\right]  \label{x15a} \\
w_{2}^{\Delta }(\tau ) &=&\nu _{2}^{\Delta }\left[ G_{M,\Delta }^{2}(\tau
)+3G_{E,\Delta }^{2}(\tau )+\frac{4\tau }{\mu _{\Delta }^{2}}G_{C,\Delta
}^{2}(\tau )\right]~,  \label{x15b}
\end{eqnarray}
where the 
magnetic, electric and Coulomb 
form factors (following~\cite{Amaro:1999be}) are taken to be:
\begin{eqnarray}
G_{M,\Delta }(\tau ) &=&2.97g_\Delta (Q^{2})  \label{x16a} \\
G_{E,\Delta }(\tau ) &=&-0.03g_\Delta (Q^{2})  \label{x16b} \\
G_{C,\Delta }(\tau ) &=&-0.15G_{M,\Delta }(\tau ),  \label{x16c}
\end{eqnarray}
with 
\begin{equation}
g_\Delta (Q^{2})\equiv \frac{G_{Ep}(\tau )}{\sqrt{1+\tau}}
\label{x17}
\end{equation}
and
\begin{equation}
G_{Ep}(\tau )=\frac{1}{\left[ 1+4.97\tau \right] ^{2}}~,  \label{x18}
\end{equation}
namely the dipole parametrization of the proton (elastic) 
electric form factor (see also Sec.~\ref{sec:cross}).

As for the quasielastic region, in the $\Delta$ domain we 
ignore terms of order $\eta_F^2$. In this approximation 
(as above, denoted by the subscript $0$) 
the RFG longitudinal and transverse $N\rightarrow\Delta$ responses will read:
\begin{eqnarray}
R_{L}^{\Delta }(\kappa ,\lambda )_{0} &=&\frac{1}{2}\Lambda_0\frac{\kappa ^{2}}{\tau 
}\left[ \left( 1+\tau \rho _{\Delta }^{2}\right) w_{2}^{\Delta }(\tau
)-w_{1}^{\Delta }(\tau )\right] \times f_{RFG}(\psi _{\Delta })  \label{x19a}
\\
R_{T}^{\Delta }(\kappa ,\lambda )_{0} &=&\frac{1}{2}\Lambda_0\left[ 2w_{1}^{\Delta
}(\tau )\right] \times f_{RFG}(\psi _{\Delta }),  \label{x19b}
\end{eqnarray}
where $\Lambda_0$ is given in Eq.~(\ref{x20b}).
As usual, for electron scattering one should add the contribution 
obtained from Eqs.~(\ref{x19a}) and (\ref{x19b}) computed with
${\cal N}=Z$ and the $p\rightarrow \Delta ^{+}$ structure functions
to the one where Eqs.~(\ref{x19a}) and (\ref{x19b}) are computed with
${\cal N}=N$ and the $n\rightarrow \Delta ^{0}$
responses. Since these processes are purely 
isovector, clearly this is equivalent to using ${\cal N}=A$ with one set of 
the structure functions. 

Again, using the guidance provided by the RFG, the above procedure is easily
generalized to the experimental response in the $\Delta$ region.  Here, as long
as density-dependent corrections ({\it i.e.,} the corrections that go as
$\eta_F^2$) are ignored as above, one should divide the experimental inclusive
electro-excitation cross section in the $\Delta$ region by
\begin{equation}
S^{\Delta }\equiv \sigma _{M}\left[ v_{L}G_{L}^{\Delta }+v_{T}G_{T}^{\Delta
}\right] ,  \label{x26}
\end{equation}
to get the scaling function $F^{\Delta }(\kappa ,\lambda )$.
By comparing Eqs.~(\ref{x19a}) and (\ref{x19b}) with 
Eqs.~(\ref{x22a}) and (\ref{x22b})
one gets for the functions $G_{L,T}^{\Delta }$ the expressions
\begin{eqnarray}
G_{L}^{\Delta } &=&\frac{\kappa }{2\tau }\left[ {\cal N}\left\{ \left( 1+\tau \rho
_{\Delta }^{2}\right) w_{2}^{\Delta }(\tau )-w_{1}^{\Delta }(\tau )\right\}
\right] +{\cal O}(\eta _{F}^{2})  \label{x27a} \\
G_{T}^{\Delta } &=&\frac{1}{\kappa }\left[ {\cal N}\left\{ w_{1}^{\Delta }(\tau
)\right\} \right] +{\cal O}(\eta _{F}^{2}).  \label{x27b}
\end{eqnarray}
Note that in Eq.~(\ref{x27b}) one has $w_{1}^{\Delta }(\tau )$, 
whereas in Eq.~(\ref{x25b}) 
the factor $\tau $ in $W_{1}(\tau )=\tau G_{M}^{2}(\tau )$ has
been taken out in front. As before, one 
should take the results for reactions
with protons with the appropriate form factors and with ${\cal N}=Z$ and add
it to the results for reactions with neutrons again with the appropriate form
factors, but now with ${\cal N}=N$. Finally, to get the superscaling function 
$f^{\Delta }(\kappa ,\lambda )$ one multiplies $F^{\Delta }(\kappa ,\lambda )$
by $k_{F}$.

With the formalism in hand, we now proceed in a manner that is analogous
to our treatment of the data in the QE region, however, now focusing on the 
$\Delta$ region. In order to isolate the contributions in the $\Delta$ region,
we have subtracted from the total experimental cross sections (with Coulomb distortion effects incorporated) the quasielastic
cross section recalculated using the universal $f^{QE}(\psi'_{QE})$ introduced
above. That is, we remove the {\em impulsive} longitudinal and transverse contributions that arise from {\em elastic} $eN$ scattering, leaving (at least) MEC effects with their associated correlations and impulsive contributions arising from {\em inelastic} $eN$ scattering. As discussed earlier, the MEC effects will be ignored in the present work as they are believed to provide relatively small corrections, and thus this yields, at least for 
$\psi'_\Delta < 0$, a
response which is largely dominated by the $\Delta$. For energy losses
beyond the maximum of the $\Delta$ peak, other resonances and, at the
larger values of $q$, the tail of deep inelastic scattering contribute. As a
consequence of the different $q$-dependencies of the various contributions, it
has not been possible using the present approach to 
analyze this $\psi'_\Delta > 0$ region further. The
region of validity of $f^\Delta(\psi'_\Delta)$ therefore will be restricted to
$\psi'_\Delta < 0$.

As shown by Eqs.~(\ref{x15a}) and (\ref{x15b}), the determination of
$f^\Delta(\kappa,\lambda)$ involves a division by a combination of the M-, E-
and C-contributions with their appropriate $q$-dependence. For the latter, we
employ the parameterizations given in Eqs.~(\ref{x16a}--\ref{x16c}) used by
Amaro {\it et al.} \cite{Amaro:1999be}. Obviously, the main contribution is
due to the M1-term, the C0 and E2 contributions to the cross sections being
minor.

\begin{figure}[hbt]
\begin{center}
  \includegraphics[scale=0.5,clip,angle=0]{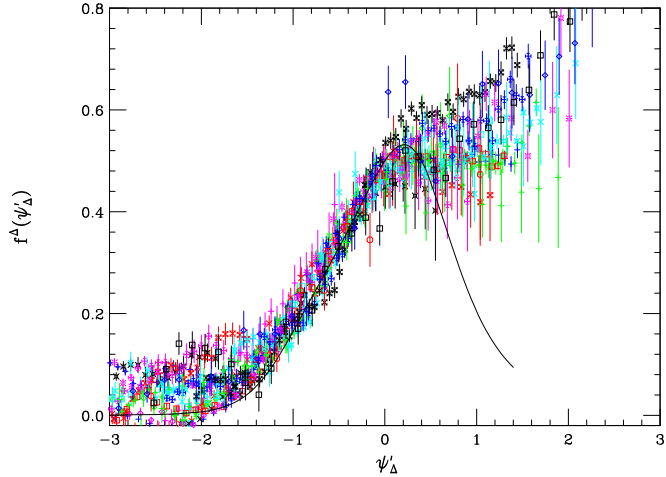}
  \parbox{13cm}{\caption[]{ Averaged experimental values of $f^\Delta (\psi'_\Delta)$
      together with a phenomenological fit (whose validity is restricted to
      $\psi^\prime_\Delta < 0$).  }\label{fdelta}}
\end{center} 
\end{figure}

In Fig.~\ref{fdelta} we show the resulting $f^\Delta(\psi'_\Delta)$ extracted
from the high-quality world data for inclusive electron scattering from
$^{12}$C and $^{16}$O in the QE and $\Delta$ regions. These data span energies extending from 300 MeV to 4 GeV and scattering angles from 12 to 145 degrees, depending on the beam energy. For this determination 
the larger-angle and higher-$q$ data are of particular importance. At small
angles and lower $q$, the $\Delta$ contribution is small, and often not present
in the available data due to limited coverage in energy-loss. As for $f^{QE}$, 
the experimental values of $f^\Delta$ have been parameterized
by a simple analytical function. We show in Fig.~\ref{fdelta} the averaged
experimental values together with this fit. As pointed out above, the validity
of the fit is restricted to $\omega$-values below the peak, {\it i.e.},
$\psi'_\Delta < 0$.

The data appear to scale reasonably well up to the peak of the $\Delta$, namely, the point where $\psi'_\Delta\cong 0$, although clearly for still higher excitation energies the scaling is broken by processes that are not well represented via $\Delta$-dominance. There is also some excess at large negative $\psi'_\Delta$ which breaks the scaling to some degree --- this is thought to be due to contributions from MEC and their associated 
correlations~\cite{Amaro02,Amaro:2001xz,Amaro03,Amaro98,DePace03,DePace04,Carlson02,Ama03,Ama94}, as discussed above. For reference we note that, for electrons of 1 GeV scattering from $^{12}$C, 
the quasielastic peak (where $\psi'_{QE}\cong 0$) occurs at $\psi'_\Delta = -1.8$, 
$-1.2$ and $-1.0$ for $\theta =$ 45, 90 and 135 degrees, respectively.
From Fig.~\ref{fpsiplaf} we see that $f^{QE}$ peaks at roughly 0.6 and thus these non-scaling contributions typically occur at the 10--15\% level in the total cross section. Below we discuss our expectations for the uncertainties incurred for our predictions of charge-changing neutrino reactions when we ignore such effects (see Sec.~\ref{sec:results}).

In passing it is important to note that in the present study we have simply taken the residual scaling function $f^\Delta$ from experiment.  While similar to $f^{QE}$ it differs in detail: it is somewhat lower, is shifted slightly and is more spread out over a wider range of scaling variable. This is perhaps not unexpected, since implicit in this approach is the fact that the $\Delta$ brings with it its own width and shift.  Only with a more microscopic model could one hope to be able to deconvolute these from the total response and see whether or not the underlying scaling function is indeed the basic $f^{QE}$ deduced above. Such an approach will be pursued in the future, although it only becomes practical when the MEC contributions are under control. For the present we limit the analysis to using two different functions $f^{QE}$ and $f^\Delta$, both deduced from phenomenological fits to electron scattering data.

\begin{figure}[hbt]
\begin{center}
  \includegraphics[scale=0.5,clip,angle=0]{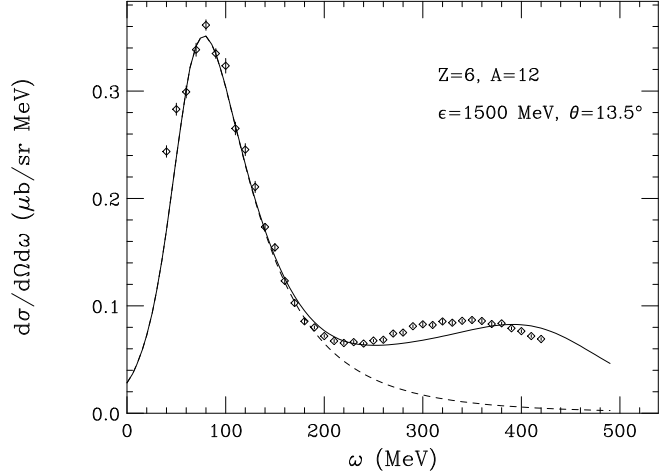}
  \parbox{13cm}{\caption[]{ Experimental $(e,e')$ cross section for $^{12}$C at 
      an incident
      electron energy of 1.5 GeV and a scattering angle of 13.5 degrees,
      together with the calculated result obtained using $f^{QE}$ and
      $f^\Delta$. The dashed curve is the QE contribution and the solid curve
      is the total including the $\Delta$.
  }\label{sigincl2}}
\end{center} 
\end{figure}

\begin{figure}[hbt]
\begin{center}
\includegraphics[scale=0.5,clip,angle=0]{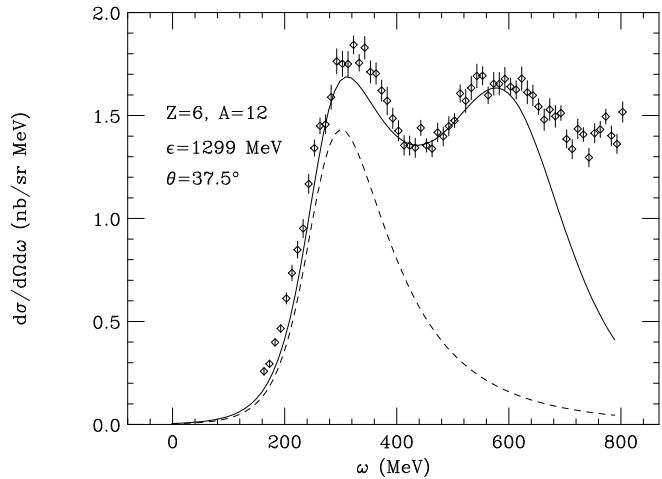}
\parbox{13cm}{\caption[]{
As for Fig.~\ref{sigincl2}, except now for an electron energy of 1.3 GeV and
 a scattering angle of 37.5 degrees.
}\label{sigincl4}} 
\end{center} 
\end{figure}

\begin{figure}[hbt]
\begin{center}
  \includegraphics[scale=0.5,clip,angle=0]{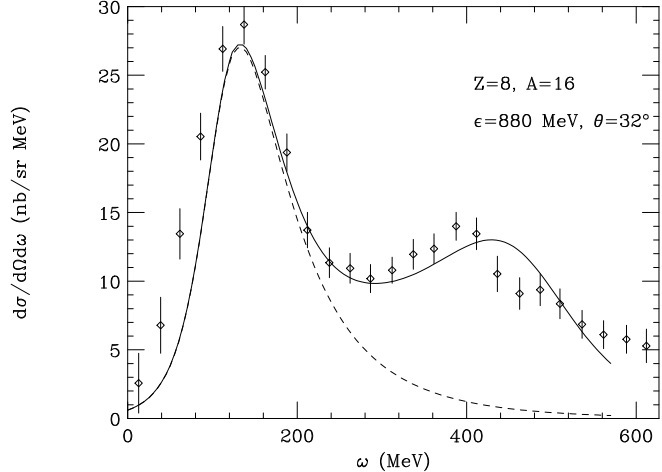}
  \parbox{13cm}{\caption[]{ Experimental $(e,e')$ cross section as above, but
      now for $^{16}$O at an incident
      electron energy of 0.88 GeV and a scattering angle of 32 degrees,
      together with the calculated result obtained using $f^{QE}$ and
      $f^\Delta$.  }\label{sigincl1}}
\end{center} 
\end{figure}

With the above ingredients, it is then possible to recalculate for every
nucleus, incident electron energy and scattering angle the inclusive cross
section for  $\omega$ below the maximum of the $\Delta$ contribution.
In order to demonstrate this, 
in Figs.~\ref{sigincl2}--\ref{sigincl1} we show
the experimental responses together with the calculated response obtained using
the parameterized $f^{QE}$ and $f^\Delta$. In particular, we have studied the accuracy of the predicted response using $(e,e')$ for
$^{12}$C and $^{16}$O and for a variety of momentum transfers, since these are the most relevant nuclei for the MiniBooNE and K2K/T2K neutrino oscillation measurements discussed in the introduction. For the data sets which do cover the
$\Delta$ region, typical deviations are 10\% or less.


\section{Charge-Changing Neutrino Reaction Formalism} 
\label{sec:ccnureac}

Among several options available (see for instance
\cite{Maieron:2003df,Kim1996,Donnelly:1978tz,Donnelly:1985,Alberico97plus})
we choose to write the charge-changing neutrino cross section in the target
lab. frame in the form
\begin{equation}
\left[ \frac{d^2 \sigma }{d\Omega dk' }\right] _{\chi }\equiv \sigma _{0}
{\cal F}_{\chi }^{2},  \label{xx10}
\end{equation}
where $\chi =+$ for neutrino-induced reactions (for example, $\nu_l+
n\rightarrow \ell^{-}+p$, where $\ell =e,\mu ,\tau $) and $\chi =-$ for
antineutrino-induced reactions (for example, $\overline{\nu_l }+p\rightarrow
\ell^{+}+n$). In Eq.~(\ref{xx10}) 
\begin{equation}
\sigma _{0}\equiv \frac{(G \cos \theta_c)^{2}}{2\pi ^{2}} 
\left[ k^{\prime }\cos 
\widetilde{\theta }/2 \right]^2 ,  \label{xx23}
\end{equation}
where $G=1.16639$x$10^{-5}$ GeV$^{-2}$ is the Fermi constant, 
$\theta_c$ is the Cabibbo angle ($\cos \theta_c=$ 0.9741)
and the generalized scattering angle $\widetilde{\theta }$ reads
\begin{equation}
\tan ^{2}\widetilde{\theta }/2\equiv \frac{|Q^{2}|}{v_{0}}, 
\label{thgen}
\end{equation}
with
\begin{equation}
v_{0}\equiv (\epsilon +\epsilon ^{\prime })^{2}-q^{2}=4\epsilon \epsilon
^{\prime }-|Q^{2}|~.  \label{xx15}
\end{equation}
Henceforth we shall assume that $m=m_\nu=0$, but will always keep $m'$ nonzero.

One additional issue arises in computing the neutrino reaction cross sections
having to do with the fact that the charged leptons in the final state are not
plane waves, but are influenced by the Coulomb potential provided by the
nucleus.  This implies that the 4-momentum of the scattered lepton,
$K'^{\mu}=(\epsilon',\mathbf{k}')$, is the local quantity, in the sense that
the 3-momentum $\mathbf{k}'$ and energy $\epsilon'=\sqrt{m'^2+k'^2}$ are
determined using the sequence of steps outlined in Sec.~\ref{sec:kine},
culminating in Eqs.~(\ref{xx7a}) and (\ref{xx7b}) for these variables.
However, the asymptotic energy-momentum is not the same as the local quantity.
Following standard procedures (see, for instance,
\cite{Alberico97plus,deForest:1966}) the Coulomb interaction can be
incorporated, at least approximately, by shifting from
$(\epsilon',\mathbf{k}')$ to asymptotic energy-momentum
$(\epsilon'_\infty,\mathbf{k}'_\infty)$:
\begin{eqnarray}
\mathbf{k}'_\infty &=& D(k') \mathbf{k}' \label{eq:Coulxx} \\
\epsilon'_\infty &=& \sqrt{m'^2+k'^2} ,
\label{eq:Coul}
\end{eqnarray} 
where 
\begin{equation}
D(k') = 1-\chi \frac{3Z\alpha}{2Rk'}
\label{eq:Coulfac}
\end{equation}
and $R\cong 1.2 A^{1/3}$ is the effective charge radius of the nucleus being
studied. Thus, our procedure is to calculate the cross sections using the 
kinematics as discussed in Sec.~\ref{sec:kine} and then at the end present
the results in terms of the asymptotic energies and momenta obtained in this
approximate manner. The only remaining issue is that the local calculations
must also be multiplied by the density-of-states factor $[D(k')]^{-1}$ to
obtain the results we present in Sec.~\ref{sec:results}.

The nuclear structure dependent quantity 
${\cal F}_{\chi }^{2}$ may be written as
\begin{equation}
{\cal F}_{\chi }^{2}=\left[ \widehat{V}_{CC}R_{CC}+2\widehat{V}%
_{CL}R_{CL}+\widehat{V}_{LL}R_{LL}+\widehat{V}_{T}R_{T}\right] +
\chi \left[
2 \widehat{V}_{T^{\prime }}R_{T^{\prime }}\right]~, 
\label{F2}
\end{equation}
that is, as a generalized Rosenbluth decomposition having Charge-Charge (CC), 
Charge-Longitudinal (CL), Longitudinal-Longitudinal (LL) and two types of Transverse 
(T,T$^\prime$) responses. 
Next we expand these response functions into their vector and
axial-vector contributions according to
\begin{eqnarray}
R_{CC} &=&R_{CC}^{VV}+R_{CC}^{AA}  \label{xy12a} \\
R_{CL} &=&R_{CL}^{VV}+R_{CL}^{AA}  \label{xy12b} \\
R_{LL} &=&R_{LL}^{VV}+R_{LL}^{AA}  \label{xy12c} \\
R_{T} &=&R_{T}^{VV}+R_{T}^{AA}  \label{xy12d} \\
R_{T^{\prime }} &=&R_{T^{\prime }}^{VA}.  \label{xy12e}
\end{eqnarray}
The lepton kinematical factors are the following:
\begin{eqnarray}
\widehat{V}_{CC} &=&1-\tan ^{2}\widetilde{\theta }/2\cdot \delta ^{2}
\label{xx13a} \\
\widehat{V}_{CL} &=&\nu +\frac{1}{\rho ^{\prime }%
}\tan ^{2}\widetilde{\theta }/2\cdot \delta ^{2}  \label{xx13b} \\
\widehat{V}_{LL} &=&\nu^2+\tan ^{2}\widetilde{%
\theta }/2\left( 1+\frac{2\nu }{\rho ^{\prime }}+\rho \cdot \delta
^{2}\right) \cdot \delta ^{2}  \label{xx13c} \\
\widehat{V}_{T} &=&\left[ \frac{1}{2}\rho +\tan ^{2}\widetilde{\theta }%
/2\right] -\frac{1}{\rho ^{\prime }}\tan ^{2}\widetilde{\theta }/2\left( 
\nu +\frac{1}{2}\rho \rho ^{\prime }\cdot \delta ^{2}\right)
\cdot \delta ^{2}  \label{xx13d} \\
\widehat{V}_{T^{\prime }} &=&\left[ \frac{1}{\rho ^{\prime }}\tan ^{2}%
\widetilde{\theta }/2\right] \left( 1-\nu \rho^{\prime } \cdot
\delta ^{2}\right) .  \label{xx13f} 
\end{eqnarray}
Here $m_\nu$ has been assumed to be zero, and the entire lepton mass
dependence occurs via the dimensionless parameter
\begin{equation}
\delta \equiv \frac{m^{\prime }}{\sqrt{|Q^{2}|}}~.  \label{xx12}
\end{equation}
Moreover, in the above we have defined
\begin{eqnarray}
\nu &\equiv & \frac{\omega}{q}=\frac{\lambda}{\kappa} \label{xx17aa} \\
\rho  &\equiv &\frac{|Q^{2}|}{q^{2}}=\frac{\tau}{\kappa^2}
=1-\nu^2  \label{xx17a} \\
\rho ^{\prime } &\equiv &\frac{q}{\epsilon +\epsilon ^{\prime }}~,
\label{xx17b}
\end{eqnarray}
which turn out to be related as follows:
\begin{equation}
\rho ^{\prime }=\frac{\tan \widetilde{\theta }/2}{\sqrt{\rho +\tan ^{2}%
\widetilde{\theta }/2}}  \label{xx18}
\end{equation}
and lie between zero and unity.
In passing we note that, using the generalized scattering angle
in Eq.~(\ref{thgen}),
\begin{eqnarray}
|Q^{2}| &=&4\epsilon \epsilon ^{\prime }\sin ^{2}\widetilde{\theta }/2
\label{xx16a} \\
v_{0} &=&4\epsilon \epsilon ^{\prime }\cos ^{2}\widetilde{\theta }/2.
\label{xx16b}
\end{eqnarray}
Finally, for use later we define
\begin{equation}
\widehat{V}_{L} \equiv \widehat{V}_{CC}-2\nu\widehat{V}_{CL}+\nu^2\widehat{V}_{LL}.
\label{eq:hatvl}
\end{equation}
For comparisons with finite-mass corrections to electron 
scattering see~\cite{Donnelly:ry}.

In the extreme relativistic limit (ERL), namely $m^{\prime}\rightarrow 0$, 
the kinematical factors are obtained by observing that in this situation
$\widetilde{\theta }$ becomes $\theta $ and 
that all of the terms containing  $\delta$ can be dropped. One then gets
\begin{eqnarray}
\widehat{V}_{CC} &\rightarrow &{v}_{CC}=1  \label{xx22a} \\
\widehat{V}_{CL} &\rightarrow &{v}_{CL}=\nu  \label{xx22b} \\
\widehat{V}_{LL} &\rightarrow &{v}_{LL}=\nu^2  \label{xx22c} \\
\widehat{V}_{L} &\rightarrow &{v}_{L}=\rho ^{2}  \label{xx22d} \\
\widehat{V}_{T} &\rightarrow &{v}_{T}=\frac{1}{2}\rho +\tan
^{2}\theta /2  \label{xx22e} \\
\widehat{V}_{T^{\prime }} &\rightarrow &{v}_{T^{\prime }}=\tan
\theta /2\sqrt{\rho +\tan ^{2}\theta /2}~,  \label{xx22f}
\end{eqnarray}
where the last three kinematical factors coincide with those employed
in electron scattering (spin observables,
coincidence electron scattering, parity-violating electron scattering, 
{\it etc.}).

In the case of the $\mu=0$ (C) and 3 (L) components of the vector current, which
is assumed to be conserved, it is possible to collapse the contributions
down to a single term: in particular, one has
\begin{eqnarray}
R_{CL}^{VV} &=& -\nu R_{CC}^{VV}
\label{xx25b} \\
R_{LL}^{VV} &=& \nu^2 R_{CC}^{VV}~.
\label{xx25c}
\end{eqnarray}
Since everything of the purely polar-vector type can be related to a single
response, traditionally we call this the longitudinal contribution, defined by
the equation $R_{L}^{VV} \equiv R_{CC}^{VV}$. Expressing the sum of the
$(\mu,\nu)=(0,0)$, $(0,3)$, $(3,0)$ and $(3,3)$ contributions, one then ends
up with a single term
\begin{equation}
\widehat{V}_{CC}R_{CC}^{VV}+2\widehat{V}_{CL}R_{CL}^{VV}+\widehat{V}%
_{LL}R_{LL}^{VV}=\widehat{V}_{L}R_{L}^{VV}\equiv X_{L}^{VV}~.  \label{xx26}
\end{equation}
This collapse into a single expression does not occur for the AA terms.
Indeed there one has
\begin{equation}
\widehat{V}_{CC}R_{CC}^{AA}+2\widehat{V}_{CL}R_{CL}^{AA}+\widehat{V}%
_{LL}R_{LL}^{AA}\equiv X_{C/L}^{AA}~.
\end{equation}
To complete the analysis one should add the two contributions
$(\mu,\nu)= (1,1)$ and $(2,2)$, which yield
\begin{equation}
\widehat{V}_{T}\left[ R_{T}^{VV}+R_{T}^{AA}\right] \equiv X_{T}~,  \label{xx30}
\end{equation}
and, as well, consider the V/A interference term where $(\mu,\nu)=(1,2)$ and
$(2,1)$,
\begin{equation}
2\widehat{V}_{T^{\prime }}R_{T^{\prime }}^{VA} \equiv  X_{T^{\prime }}~.
\label{xx32b}
\end{equation}
The full response will then be (see Eq.~(\ref{F2})) 
\begin{equation}
{\cal F}_{\chi }^{2}=X_{L}^{VV}+X_{C/L}^{AA}+X_{T}+ \chi X_{T^{\prime }}.
\label{F2X}
\end{equation}

\subsection{Single-nucleon responses in the QE region}
\label{sec:snqe}

The single-nucleon responses in the QE region all begin with the basic
vector and axial-vector currents involving $N\to N$ matrix elements:
\begin{eqnarray}
{j}_{V}^{\mu } &=& {\bar u}(P') \left[ F_1\gamma ^{\mu }+\frac{i}{2m_{N}}%
F_2 \sigma^{\mu\nu}Q_{\nu }\right] u(P),  \label{xx28xx} \\
{j}_{A}^{\mu } &=& {\bar u}(P') \left[ G_{A}\gamma ^{\mu }+\frac{1}{2m_{N}}%
G_{P}Q^{\mu }\right] \gamma ^{5} u(P),  \label{xx28} 
\end{eqnarray}
with $Q^\mu=P'^\mu - P^\mu$. Indeed, in the scaling analyses discussed in
Sec.~\ref{sec:scalqe}, the former was used together with the usual
relationship between the Dirac/Pauli form factors and the Sachs form factors,
$G_E=F_1-\tau F_2$ and $G_M=F_1+F_2$, to obtain expressions such as those in
Eqs.~(\ref{y23a},\ref{y23b}).  The total current is then
$j^\mu=j^\mu_V-j^\mu_A$.

In fact, for the purely polar-vector contributions we have 
\begin{eqnarray}
R_{L}^{VV} &=&\frac{1}{\rho }\left[ G_{E}^{(1)}\right] ^{2}
  \label{xxx25a} \\
R_{T}^{VV} &=&2\tau \left[ G_{M}^{(1)}\right] ^{2}~,  \label{xx31a}
\label{xxx25b}
\end{eqnarray}
which were used above. Here $G_{E,M}^{(1)}$ are the nucleon's EM isovector
form factors.  We re-emphasize that, as written, these results contain effects
from the motion of the nucleons in the nucleus up to first-order in $\eta_F$
and only terms of order $\eta_F^2$ and beyond have been neglected.

For the axial-vector current proceeding from Eq.~(\ref{xx28}) one has
\begin{eqnarray}
R_{CC}^{AA} &=& \frac{\nu ^{2}}{\rho }
 \left[ G_{A}^{\prime \,(1)}\right] ^{2}  \label{xx27a} \\
R_{CL}^{AA} &=& -\frac{\nu}{\rho }
 \left[ G_{A}^{\prime \,(1)}\right] ^{2}  \label{xx27b} \\
R_{LL}^{AA} &=& \frac{1}{\rho }
 \left[ G_{A}^{\prime \,(1)}\right] ^{2}  \label{xx27c} \\
R_{T}^{AA} &=&2(1+\tau )\left[ G_{A}^{(1)}\right] ^{2}~,  \label{xx31b}
\end{eqnarray}
defining the following combination of axial-vector and induced 
pseudoscalar form factors 
\begin{equation}
G_A^{\prime \,(1)}(\tau)= G_A^{(1)}(\tau)-\tau G_P^{(1)}(\tau)~,  \label{xx29}
\end{equation}
namely, the axial-vector analog of the relationship between the
Dirac, Pauli and Sachs form factors. 
It is of importance to realize that in the ERL 
(see Eqs.~(\ref{xx22a}-\ref{xx22c}))  
$X_{C/L}^{AA}\rightarrow 0$: hence this term crucially depends upon
the final lepton mass, yielding for the C/L single-nucleon QE response
\begin{equation}
X_{C/L}^{AA} = \tan ^{2}\widetilde{\theta }/2 
 \left[ G_{A}^{\prime \,(1)}\right] ^{2} \cdot (1+\delta^2)\delta^2 .
\label{xx29xx}
\end{equation}
Finally the V/A interference term is
\begin{equation}
R_{T^{\prime }}^{VA} = 2\sqrt{\tau (1+\tau )}G_{M}^{(1)}G_{A}^{(1)}~.
\label{xx32a} 
\end{equation}

\subsection{Single-nucleon responses in the $\Delta$ region}
\label{sec:sndel}

In this subsection we consider the elementary reactions
\ba
\nu_\mu p &\to& \mu^-\Delta^{++}
\label{nup}
\\
\nu_\mu n &\to& \mu^- \Delta^{+}
\label{nun}
\\
\bar\nu_\mu p &\to& \mu^+\Delta^{0} 
\label{nubarp}
\\
\bar\nu_\mu n &\to& \mu^+\Delta^{-}~.
\label{nubarn}
\ea
The associated currents are~\cite{Alvarez-Ruso:1998hi} 
\be
J^\mu(q)={\cal T} 
\bar u_\alpha^{(\Delta)}(p',s')\Gamma^{\alpha\mu} u(p,s),
\ee
where the isospin factor ${\cal T}$ is $\sqrt{3}$ for $\Delta^{++}$ 
and $\Delta^{-}$ 
production and $1$ for $\Delta^{+}$ and $\Delta^{0}$ production, 
$u_\alpha^{(\Delta)}(p',s')$ and $u(p,s)$ are the Rarita-Schwinger and 
Dirac spinors for a $\Delta$ and a nucleon with momenta $p'=p+q$ and $p$ 
and spin $s'$ and $s$, respectively. 
For the vertex tensor we take~\cite{Alvarez-Ruso:1998hi}

\ba
\Gamma^{\alpha\mu} &=& 
\left[
\frac{C_3^V}{m_N} \left(g^{\alpha\mu} \sla q - q^\alpha \gamma^\mu \right)
+\frac{C_4^V}{m_N^2} \left(g^{\alpha\mu} q\cdot p'-q^\alpha p'^\mu\right)
+\frac{C_5^V}{m_N^2} \left(g^{\alpha\mu} q\cdot p-q^\alpha p^\mu\right)
\right] \gamma_5
\nonumber\\
&+&
\left[
\frac{C_3^A}{m_N} \left(g^{\alpha\mu} \sla q - q^\alpha \gamma^\mu \right)
+\frac{C_4^A}{m_N^2} \left(g^{\alpha\mu} q\cdot p'-q^\alpha p'^\mu\right)
+C_5^A g^{\alpha\mu} 
+\frac{C_6^A}{m_N^2} q^\alpha q^\mu\right]~. \nonumber\\
&&
\ea
We recall that CVC implies
$C_6^V=0$ and PCAC yields $C_6^A=C_5^A (\mu_\pi^2+4\tau)^{-1}$, with 
$\mu_\pi=m_\pi/m_N$, $m_\pi$ being the pion mass.

The hadronic tensor
\be
w^{\mu\nu} = \mu_\Delta \Tr\left\{J^{\mu\dagger} (q) 
J^\nu (q)\right\},
\label{Tr}
\ee
with $\mu_\Delta=m_\Delta/m_N$ as above, can be rewritten in the form
\be
w^{\mu\nu} = {\cal T}^2 \mu_\Delta \Tr\left\{P_{\beta\alpha}(p')
\left(\gamma_0 \Gamma^{\dagger\alpha\mu}\gamma_0\right)
\Lambda(p) \Gamma^{\beta\nu} \right\}\,,
\ee
where 
\ba
P_{\beta\alpha}(p') &=& 
\sum_{s'} u_\beta^{(\Delta)}(p',s')\bar u_\alpha^{(\Delta)}(p',s')
\\
&=&-\frac{\sla p'+m_\Delta}{2 m_\Delta}
\left(g_{\beta\alpha}-\frac{2}{3}\frac{p'_\beta p'_\alpha}{m_\Delta^2}
+\frac{1}{3}\frac{p'_\beta\gamma_\alpha-p'_\alpha\gamma_\beta}{m_\Delta}
-\frac{1}{3}\gamma_\beta\gamma_\alpha\right)
\ea
is the Rarita-Schwinger projector, while as usual
\be
\Lambda(p)=\frac{\sla p+m_N}{2 m_N}
\ee
is the nucleon projector.

A lengthy calculation then yields
\ba
 w^{\mu\nu} &=& 
w^{\mu\nu}_{VV}+w^{\mu\nu}_{AA}+w^{\mu\nu}_{VA}
\ea
where
 \ba
w^{\mu\nu}_{VV} &=& 
-w_{1V}(g^{\mu\nu}+\frac{\kappa^\mu\kappa^\nu}{\tau})
+w_{2V}(\eta^\mu+\rho_\Delta\kappa^\mu)(\eta^\nu+\rho_\Delta\kappa^\nu)~,
\label{eq:wmunuVV}
\\
w^{\mu\nu}_{AA} &=& 
-w_{1A}(g^{\mu\nu}+\frac{\kappa^\mu\kappa^\nu}{\tau})
+w_{2A}(\eta^\mu+\rho_\Delta\kappa^\mu)(\eta^\nu+\rho_\Delta\kappa^\nu)
\nonumber\\
&&-u_{1A}\frac{\kappa^\mu\kappa^\nu}{\tau}
+u_{2A}(\kappa^\mu\eta^\nu+\eta^\mu\kappa^\nu)
\label{eq:wmunuAA}
\ea
and  
\begin{equation}
w^{\mu\nu}_{VA} =  
2 i w_3 \epsilon^{\alpha\beta\mu\nu} \eta_\alpha\kappa_\beta
\label{eq:wmunuVA}
\end{equation}
are the Vector-Vector, Axial-Axial and Vector-Axial interference
hadronic tensors, respectively.
The functions $u_{1A}$ and $u_{2A}$ reflect the non-conservation of
the axial-vector current and will be discussed in the Appendix.
The functions $w_i$ entering above are obtained
by performing the traces in Eq.~(\ref{Tr}) and using the 
on-shell conditions $\eta^2=1$
and $\eta\cdot\kappa=\tau\rho_\Delta=\tau+\frac{1}{4}(\mu_\Delta^2-1)$.
The results are collected in the Appendix.

\subsection{Nuclear cross sections and response functions}
\label{sec:cross}

To compute the cross section in Eq.~(\ref{xx10}), the factor 
${\cal F}^2_\chi$ in Eq.~(\ref{F2X}) is required, and thus the 
nuclear response functions $R_i$ are needed.
The last can be expressed in terms of the nuclear tensor $W^{\mu\nu}$
according to
\ba
R_L^{VV} &=& W^{00}_{VV}
\\
R_{CC}^{AA} &=& W^{00}_{AA}
\\
R_{CL}^{AA} &=& -W^{03}_{AA}
\\
R_{LL}^{AA} &=& W^{33}_{AA}
\\
R_{T}^{VV} &=& W^{11}_{VV}+W^{22}_{VV}
\\
R_{T}^{AA} &=& W^{11}_{AA}+W^{22}_{AA}
\\
R_{T'}^{VA} &=& -i W^{12}_{VA}~.
\label{tprime}
\ea

Since in the QE domain the nuclear tensor is well-known, here we focus on the
$\Delta$ sector, where using Eq.~(\ref{x20b}) $W^{\mu\nu}$ 
reads~\cite{Amaro:1999be}
\ba
\left[W^{\mu\nu}\right]^{\Delta} &=&
\frac{3{\cal N}}{8 m_N\eta_F^3\kappa} 
\int_{\epsilon_0}^{\epsilon_F}w^{\mu\nu}(\epsilon) 
\theta(\epsilon_F-\epsilon_0)
d\epsilon
\nonumber\\
&=& \frac{1}{2} \Lambda_0 f_{RFG}(\psi_\Delta) 
U^{\mu\nu},
\label{WmunuD}
\ea
where ${\cal N}=Z$ for reactions on protons, (\ref{nup}) and (\ref{nubarp}),
whereas ${\cal N}=N$ for reactions on neutrons, (\ref{nun}) and (\ref{nubarn}). 
Again, only isovector form factors enter.

As discussed previously (see Sec.~\ref{sec:scaldel}), 
the RFG superscaling function in the $\Delta$ domain is given by
\be
f_{RFG}(\psi_\Delta)=\frac{3}{4}(1-\psi_\Delta^2)\theta(1-\psi_\Delta^2)
\ee
and
\be
U^{\mu\nu}=\frac{1}{\epsilon_F-\epsilon_0}
\int_{\epsilon_0}^{\epsilon_F}w^{\mu\nu}(\epsilon) d\epsilon\,,
\label{eq:Umunu}
\ee
where
\be
\epsilon_0=\kappa \sqrt{\rho_{\Delta}^{2}+1/\tau }-\lambda \rho _{\Delta}
\ee
represents the minimum energy a nucleon should have in order to take
part into the $\Delta$ electro-excitation process and
$\epsilon_F=\sqrt{1+\eta_F^2}$ is the Fermi energy. Note that, from Eq.~(\ref{x3}), one has
\begin{equation}
\psi _{\Delta}= \left[ \frac{1}{\xi _{F}}\left( \epsilon_0-1\right) \right] ^{1/2}\times \left\{ 
\begin{array}{cc}
+1 & \lambda \geq \lambda _{\Delta}^{0} \\ 
-1 & \lambda \leq \lambda _{\Delta}^{0}
\end{array}
\right. ,  \label{x3delta}
\end{equation}
with $\lambda _{\Delta}^{0}=\frac{1}{2}\left[ \sqrt{\mu _{\Delta}^{2}+4\kappa ^{2}}-1\right]$ using Eq.~(\ref{x5}).

The required components of the $U^{\mu\nu}$ tensor can be explicitly computed
using Eqs.~(\ref{eq:wmunuVV}-\ref{eq:wmunuVA}) and performing the integral 
in Eq.~(\ref{WmunuD}). They turn out to be
\ba
U^{00}_{VV} &=& \frac{\kappa^2}{\tau}
\left[-w_{1V}(\tau)+(1+\tau\rho_\Delta^2) w_{2V}(\tau) +{\cal
    D}(\kappa,\tau)w_{2V}(\tau) \right]
\nonumber\\
&&\label{U00VV}
\ea
\ba
U^{00}_{AA}
&=& \frac{\kappa^2}{\tau} \left[-w_{1A}(\tau)+(1+\tau\rho_\Delta^2)
w_{2A}(\tau)+{\cal D}(\kappa,\tau)w_{2A}(\tau)
\right]
\nonumber\\
&-&
\frac{\lambda^2}{\tau} u_{1A}(\tau)
+\lambda(\epsilon_F+\epsilon_0)
u_{2A}(\tau)
\label{U00AA}
\ea
\ba
U^{03}_{AA}
&=& \left(\frac{\lambda}{\kappa}\right)
\Bigg\{
\frac{\kappa^2}{\tau} \left[-w_{1A}(\tau)+
(1+\tau\rho_\Delta^2)w_{2A}(\tau)+{\cal D}(\kappa,\tau)w_{2A}(\tau)\right]
\nonumber\\
&-&
\frac{\kappa^2}{\tau} u_{1A}(\tau)
+\left[\frac{\lambda^2+\kappa^2}{2\lambda}(\epsilon_F+\epsilon_0)
-\tau\rho_\Delta\right]
u_{2A}(\tau)\Bigg\}
\label{U03AA}
\ea
\ba
U^{33}_{AA}
&=& \left(\frac{\lambda}{\kappa}\right)^2
\Bigg\{
\frac{\kappa^2}{\tau} \left[-w_{1A}(\tau)+
(1+\tau\rho_\Delta^2)w_{2A}(\tau)+{\cal D}(\kappa,\tau)w_{2A}(\tau)\right]
\nonumber\\
&-&\frac{\kappa^4}{\tau\lambda^2}u_{1A}(\tau)
+\frac{\kappa^2}{\lambda^2}\left[\lambda(\epsilon_F+\epsilon_0)
-2\tau\rho_\Delta\right]u_{2A}(\tau)\Bigg\}
\label{U33AA}
\ea
for pieces of the tensor having $\mu$ or $\nu$ equal to $0$ or $3$, while
for transverse projections one has
\ba
U^{11}_{VV}+U^{22}_{VV} 
&=&
2 w_{1V}(\tau)+{\cal D} (\kappa,\tau) w_{2V}(\tau)
\\
U^{11}_{AA}+U^{22}_{AA} 
&=&
2 w_{1A}(\tau)+{\cal D} (\kappa,\tau) w_{2A}(\tau)
\\
U^{12}_{VA} 
&=&
2 i\sqrt{\tau(1+\tau\rho_\Delta^2)}\left[1+{\cal D}'(\kappa,\tau)\right] 
w_{3}(\tau),
\ea
where 
\ba
{\cal D}(\kappa,\tau) &=& \frac{\tau}{\kappa^2}
\left[\frac{1}{3}\left(\epsilon_F^2+\epsilon_0\epsilon_F
+\epsilon_0^2\right)+\lambda\rho_\Delta\left(\epsilon_F+\epsilon_0\right)
+\lambda^2\rho_\Delta^2 \right]-1-\tau\rho_\Delta^2
\label{calD}
\\
{\cal D}'(\kappa,\tau) 
&=& \frac{1}{\kappa}\sqrt{\frac{\tau}{1+\tau\rho_\Delta^2}}
\left[\lambda\rho_\Delta+\frac{1}{2}\left(\epsilon_F+\epsilon_0\right)
\right]-1~.
\ea
%In the above, in the present circumstances the terms involving 
%${\cal D}$ and ${\cal D}'$, which reflect the impact of the
%medium on the single-nucleon $\Delta$ electro-excitation, go as
%$\eta_F^2$ and therefore can be disregarded
%without appreciable consequences on the results, as noted previously.

Thus we are now in a position to assemble the various factors and provide
predictions for the charge-changing neutrino reaction cross sections. The
single-nucleon cross sections in both the QE and $\Delta$ regions are given
above and, from the scaling analysis presented in Sec.~\ref{sec:supers}, we
have experimentally determined scaling functions $f^{QE}(\psi'_{QE})$ and $f^\Delta(\psi'_\Delta)$. As usual, we have used scaling variables 
$\psi'_{QE}$ and $\psi'_\Delta$ in which we have included a small energy 
shift by replacing $\omega$ with $\omega-E_{shift}\equiv \omega'$,
$\lambda\to\lambda'$, {\it etc.} For charge-changing neutrino
reactions one proceeds to analogous states in the neighboring nuclei, which
are shifted from their positions in the target nucleus at least by the Coulomb
energy.  Accordingly, while a very small effect, in this work we have modified
$E_{shift}$ from its value as obtained in studies of electron scattering by
these additional amounts. In the case of mass-12 this means that upon
comparing the excitation energy of the analog of the nitrogen and boron ground
states ({\it i.e.}, the energy 15.110 MeV of the lowest $J^{\pi}T=1^+1$ state in
carbon), we should add $16.827-15.110=1.717$ MeV when making transitions to
nitrogen (neutrino reactions) and subtract $15.110-13.880=1.230$ MeV when
making transitions to boron (antineutrino reactions) from the canonical value
of $E_{shift}=20$ MeV. These are then used in the definitions of $\psi'$ in
each case. Note that we use the correct ground state masses of the three
nuclei in establishing the lepton scattering kinematics discussed in
Sec.~\ref{sec:kine}.

Finally, in computing the results to be presented in the next section the
following form factors have been employed. 
For the vector sector we use the H\"{o}hler parameterization 8.2
\cite{Hoehler:1976}
%\ba
%G_D^V(\tau)&=&(1+\lambda_D^V\tau)^{-2}\\
%\eta(\tau)&=&(1+\lambda_\eta\tau)^{-1}\\
%f_{1p}(\tau)&=&\frac{1+\tau(1+\lambda_{p})}{1+\tau} G_D^V(\tau)\\
%f_{2p}(\tau)&=&\frac{\lambda_{p}}{1+\tau} G_D^V(\tau)\\
%f_{1n}(\tau)&=&\frac{\tau\lambda_{n}[1-\eta(\tau)]}{1+\tau} G_D^V(\tau)\\
%f_{2n}(\tau)&=&\frac{\lambda_{n}[1+\tau\eta(\tau)]}{1+\tau} G_D^V(\tau)\\
%F_1(\tau)&=& f_{1p}(\tau)-f_{1n}(\tau)\\
%F_2(\tau)&=& f_{2p}(\tau)-f_{2n}(\tau)\\
%G_E^{(1)}(\tau)&=& F_1(\tau)-\tau F_2(\tau)\\
%G_M^{(1)}(\tau)&=& F_1(\tau)+ F_2(\tau) \ea with $\lambda_D^V=4.97$,
%$\lambda_\eta=5.6$, $\lambda_p= 1.793$ and $\lambda_n= -1.913$, 
and for the axial-vector sector we use
 \ba
G_D^A(\tau)&=&(1+\lambda_D^A \tau)^{-2}\\
G_A^{(1)}(\tau)&=& g_A G_D^A(\tau)\\
G_P^{(1)}(\tau)&=&\frac{1}{1/\lambda'_A +\tau} G_A^{(1)}(\tau)
\ea 
with
$\lambda_D^A=3.32$ (corresponding to axial-vector mass $M_A=1032$ MeV),
$\lambda'_A=(2m_N/m_\pi)^2=180$ and $g_A=1.26$.

\section{Neutrino Cross Section Predictions}
\label{sec:results}

\begin{figure}[hbt]
\begin{center}
\hspace*{1.5mm}\includegraphics[scale=0.5,clip]{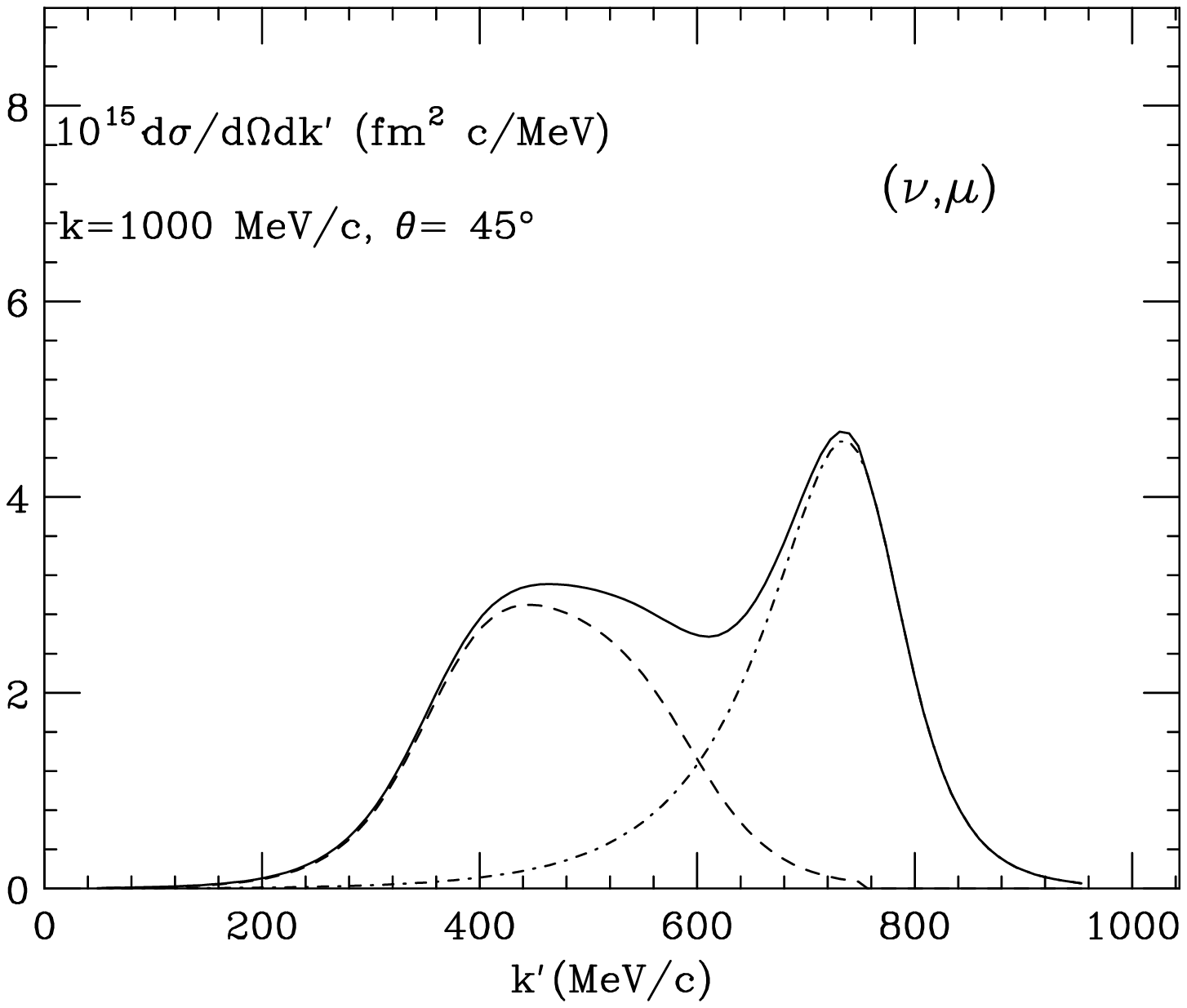}\includegraphics[scale=0.5,clip]{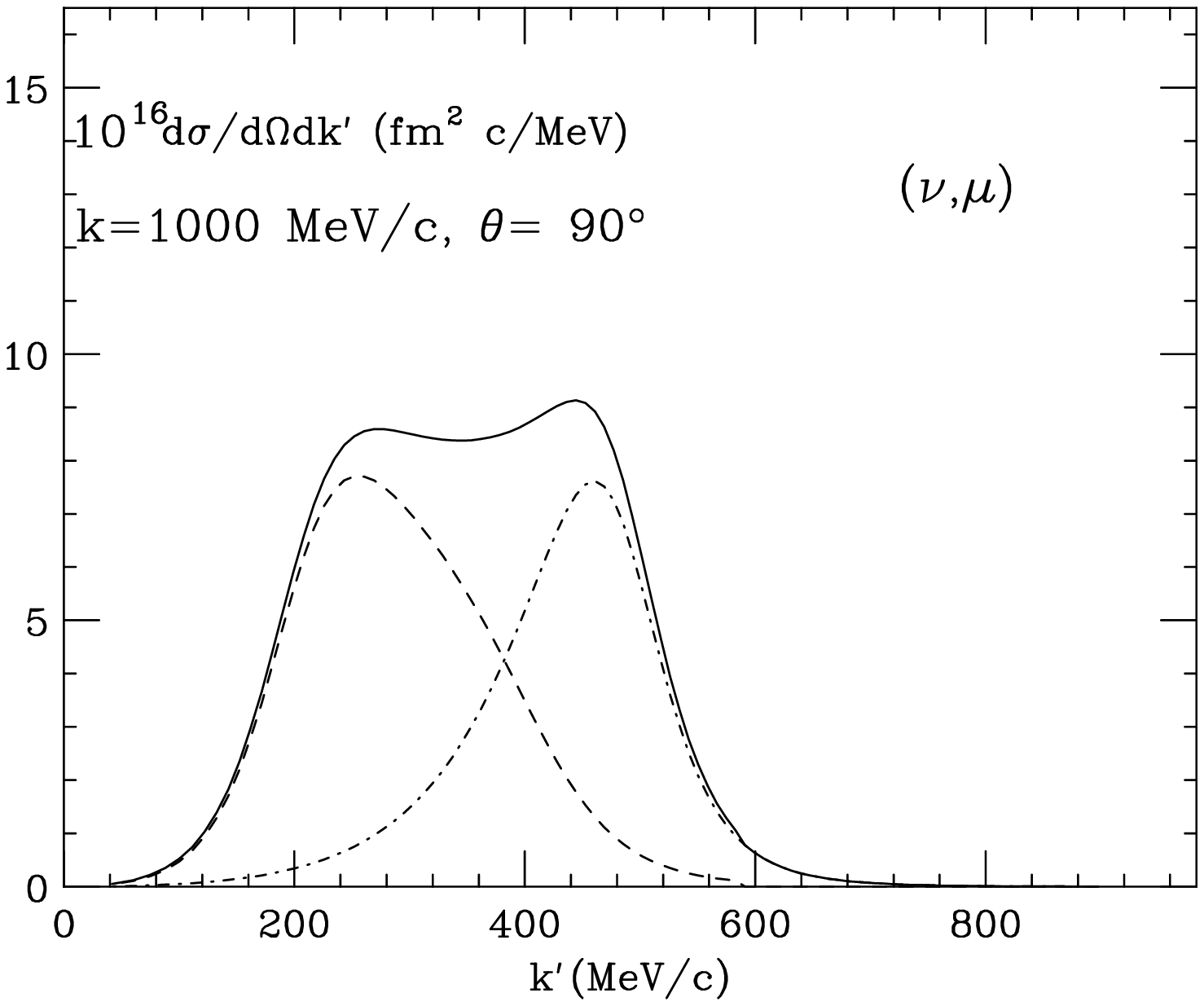}
\includegraphics[scale=0.5,clip]{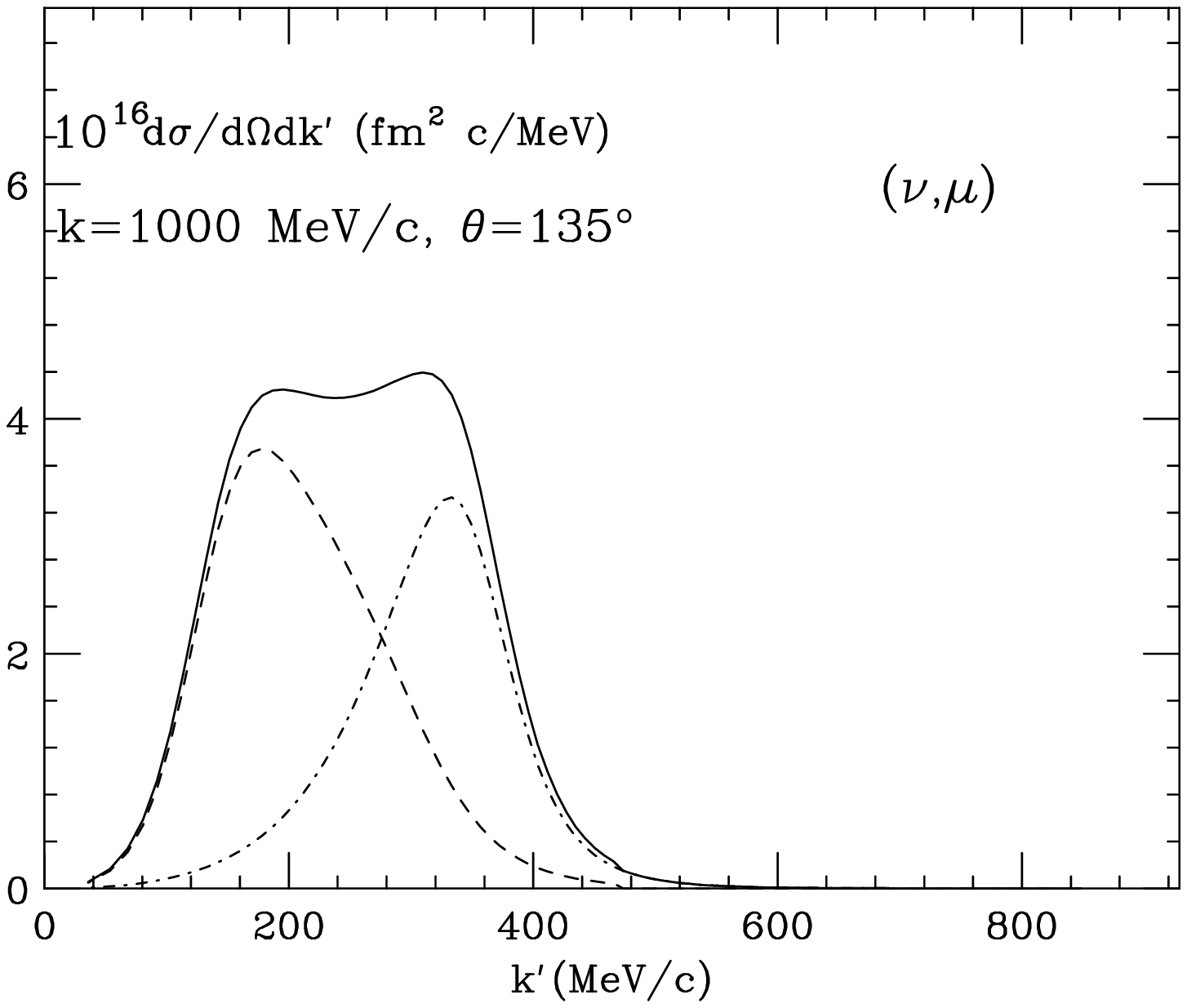}\includegraphics[scale=0.5,clip]{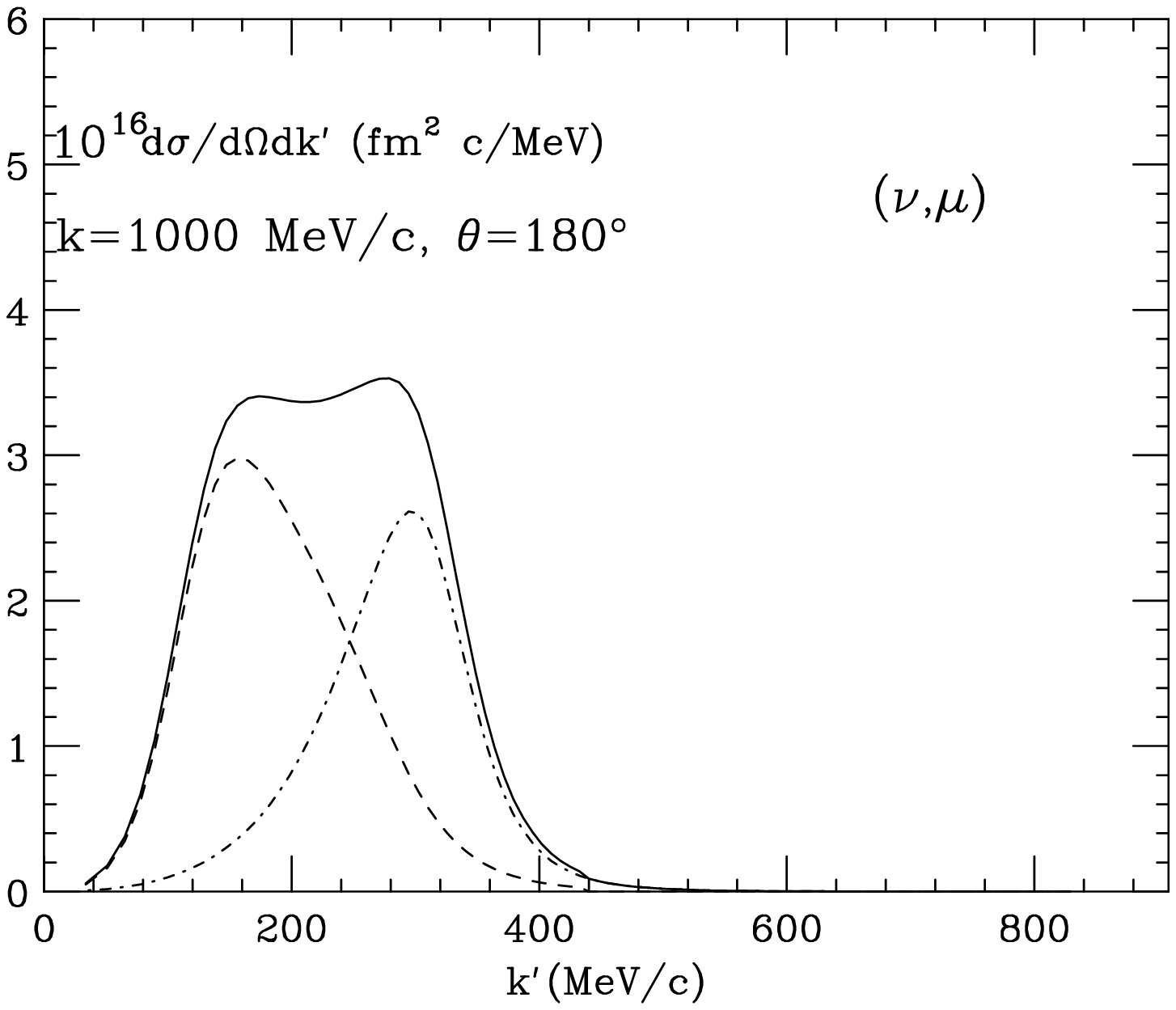}
\parbox{13cm}{\caption[]{Charge-changing neutrino reactions $(\nu_\mu,\mu^-)$ on $^{12}$C
for 1 GeV neutrinos and neutrino-muon scattering angles of 45, 90, 135 and 180 degrees. 
The cross sections are plotted versus the final-state muon momentum $k'$. The 
dash-dotted curves give the QE contribution, the dashed curves the $\Delta$ 
contribution and the solid curves the total. As discussed in the text, results 
for values of $k'$ lying below the $\Delta$ peak (higher excitation energies than that
of the $\Delta$) must be viewed with caution.
}\label{neutoncarbon}} 
\end{center} 
\end{figure}

\begin{figure}[hbt]
\begin{center}
\includegraphics[scale=0.5,clip]{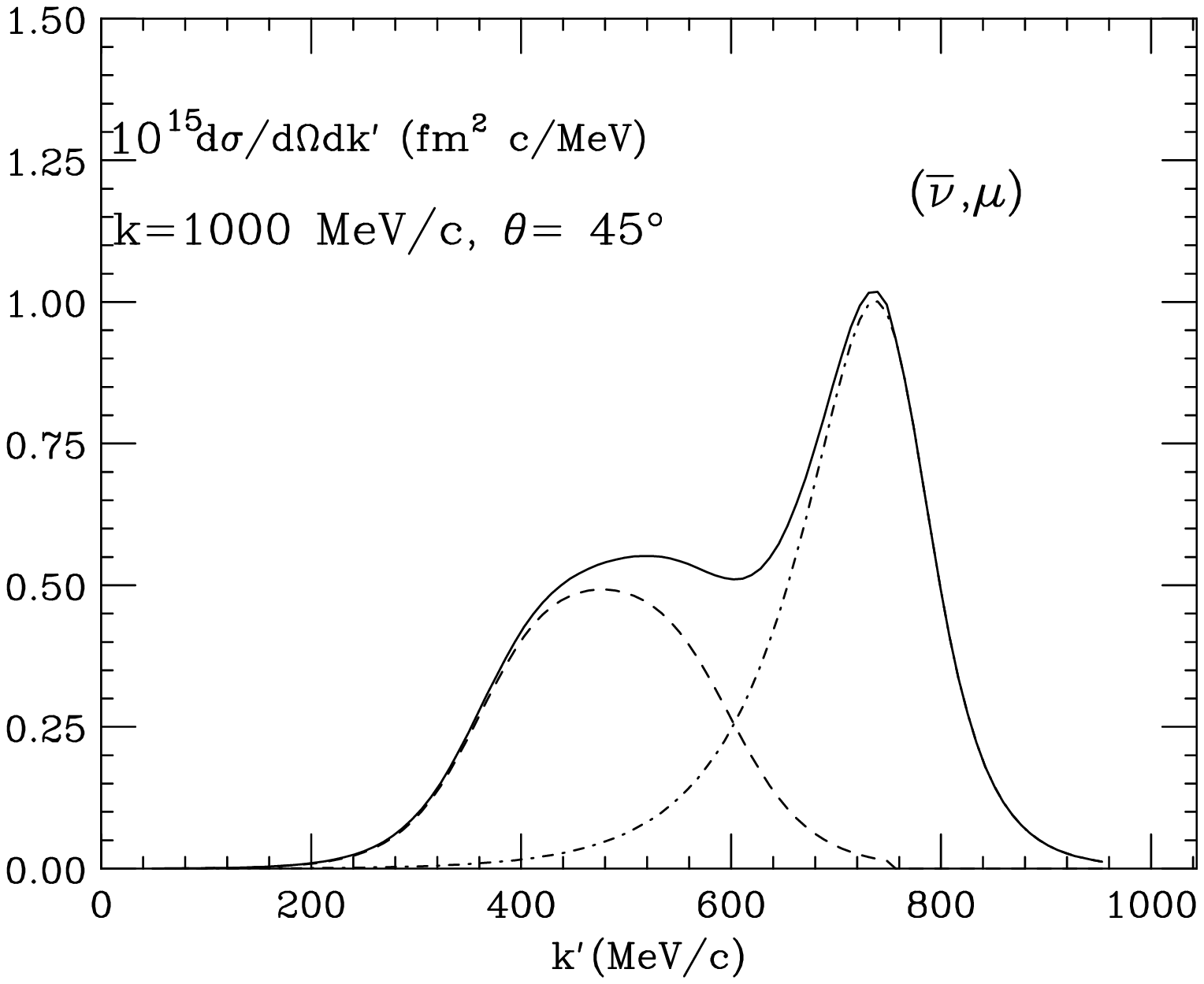}\includegraphics[scale=0.5,clip]{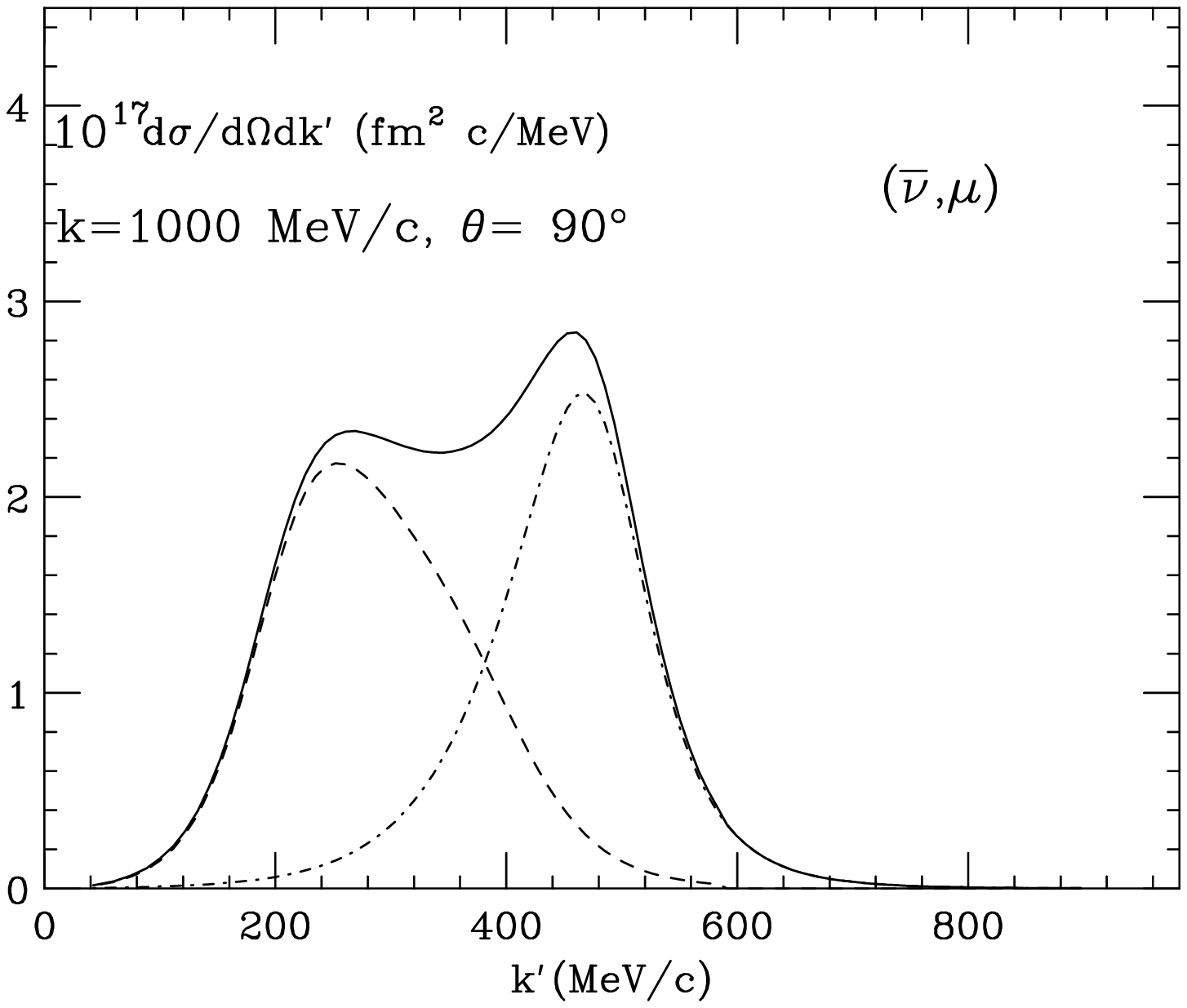}
\includegraphics[scale=0.5,clip]{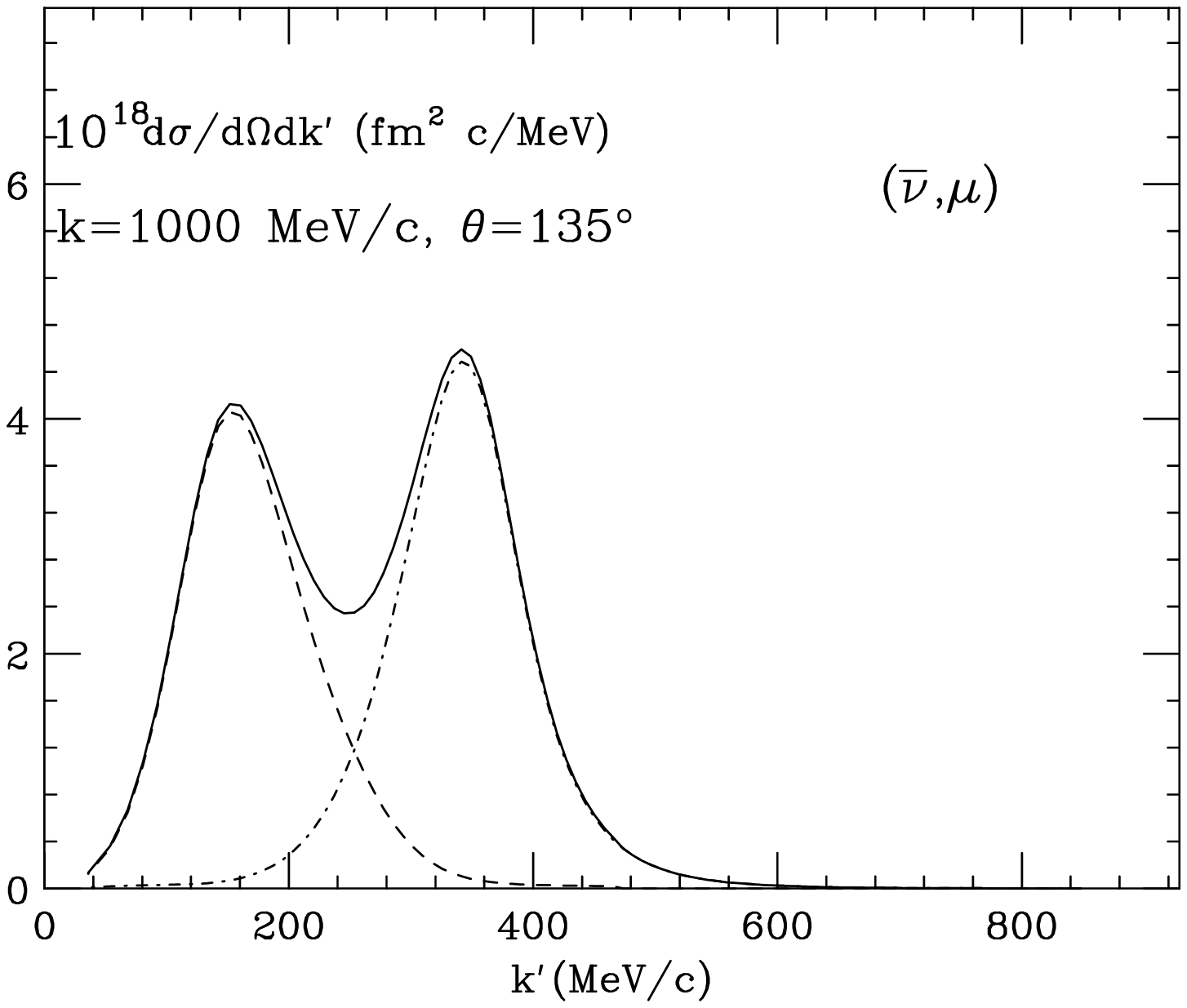}\includegraphics[scale=0.5,clip]{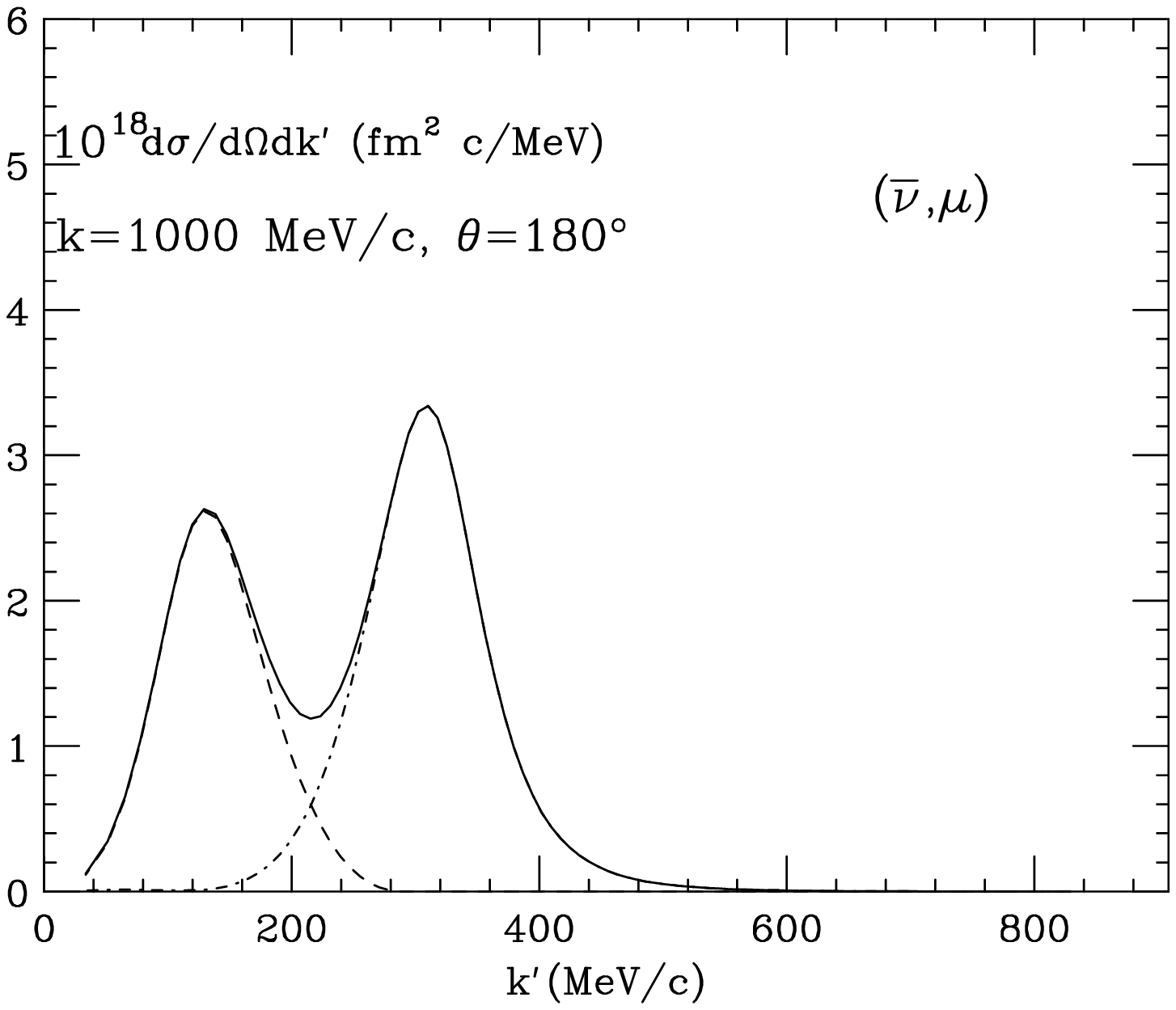}
\parbox{13cm}{\caption[]{As for Fig.~\ref{neutoncarbon}, but now for antineutrino 
reactions $({\bar{\nu}}_\mu,\mu^+)$.
}\label{antineutoncarbon}} 
\end{center} 
\end{figure}

\begin{figure}[hbt]
\begin{center}
\hspace*{-1cm}\includegraphics[scale=0.5,clip]{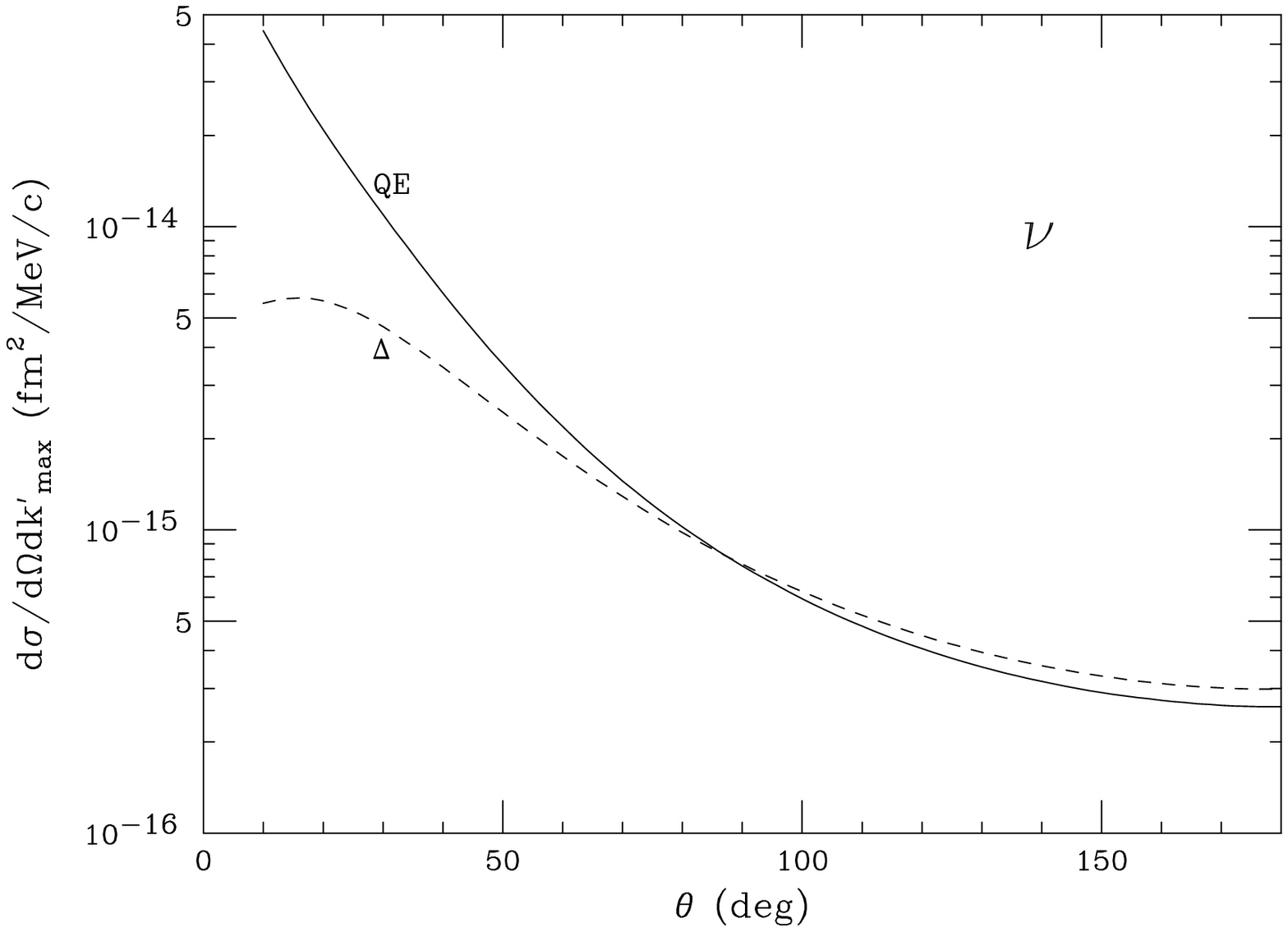}\includegraphics[scale=0.5,clip]{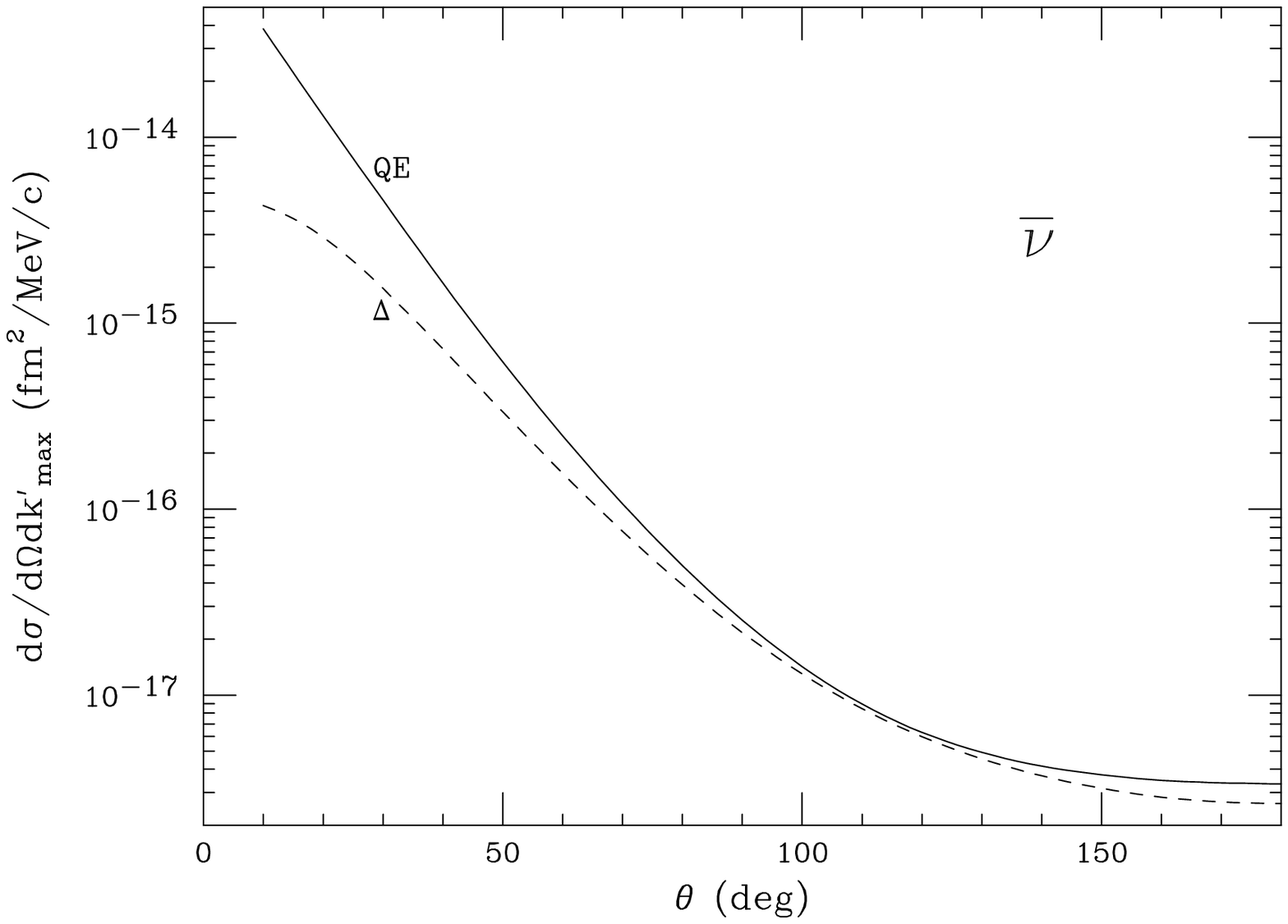}
\parbox{13cm}{\caption[]{Angular distributions for the results in 
Figs.~\ref{neutoncarbon} and \ref{antineutoncarbon}
at the tops of the QE and $\Delta$ peaks. Since the neutrino or antineutrino energy
is fixed to 1 GeV and the kinematics are chosen to be at $\psi^\prime_{QE}=0$ 
and $\psi^\prime_\Delta=0$, the muon momentum $k^\prime$ must vary with $\theta$.
}\label{angdistns}} 
\end{center} 
\end{figure}

The inclusive charge-changing neutrino reaction cross sections $^{12}$C$(\nu_\mu,\mu^-)$ and $^{12}$C$({\bar{\nu}}_\mu,\mu^+)$ that result from the scaling analyses presented above are shown in Figs.~\ref{neutoncarbon} and \ref{antineutoncarbon}, respectively. A neutrino or antineutrino energy of 1 GeV has been selected as being representative of kinematics where the scaling approach should be expected to work well and a selection of scattering angles (between incident neutrino or antineutrino and produced charged muon) has been made. Note that, since the predictions shown here are given as functions of the muon momentum $k'$, the QE peak lies to the right ({\it i.e.}, lower excitation energy) of the $\Delta$ peak (higher excitation energy). Note that the Coulomb distortion correction from Eq.~(\ref{eq:Coulxx}) has been included here, although for simplicity we have labelled the figures using
$k'$ rather than $k'_\infty$. 
As discussed above, the predictions at momenta to the left of the $\Delta$ peak (excitations lying above the $\Delta$ region) are unreliable, since our scaling approach does not fully account for meson production, including resonances other than the $\Delta$, and DIS processes. 

Corresponding angular distributions are shown in Fig.~\ref{angdistns} for kinematics chosen to lie at the peaks of the QE (solid curves) and $\Delta$ (dashed curves).

One striking feature of the results is the dramatic differences seen in comparing neutrino and antineutrino cross sections at backward angles. The latter are typically about two orders of magnitude smaller than the former under those conditions. This is due to the fact that the transverse (vector and axial-vector) contribution to both the QE and $\Delta$ responses is accidentally roughly the same in magnitude as the one arising from the V/A interference for large scattering angles at the chosen energy. Specifically, for instance in the QE case, one has $X_T$ roughly equal
to $X_{T'}$ (see Eqs.~(\ref{xx30}), (\ref{xx32b}), (\ref{xx31a}), (\ref{xx31b}) 
and (\ref{xx32a})). 
For neutrinos these constructively interfere, whereas for antineutrinos they 
tend to cancel and produce much reduced cross sections. Indeed, the cancellation 
is so severe that the VV(CC, CL and LL) and AA(CC, CL and LL) terms can yield 
significant contributions to the total cross section. In the QE region these VV
terms yield as much as 1/3 of the cross section, while the AA terms are negligible.
In contrast, in the $\Delta$ region the reverse is true, with the AA(CC, CL and LL)
terms even providing the majority of the cross section at large angles in this case. 
Small changes in the model (for example, the inclusion of MEC) could have very large effects on the predictions for antineutrinos and hence the results shown in this case, especially for large scattering angles, should be viewed with caution --- the important observation is that the antineutrino cross sections are predicted to be strongly suppressed due to the accidental cancellation.

For neutrinos, where no such strong cancellation occurs, the cross sections are typically dominated by the VV(T), AA(T) and VA(T$^\prime$) contributions. In the QE region the VV(CC, CL and LL) pieces contribute only about 5\% at 45 degrees and fall to negligible corrections at backward angles, whereas the AA(CC, CL and LL)  contributions are negligible for all angles considered. In the $\Delta$ regions the converse is true: the AA(CC, CL and LL) pieces contribute about 5\% or less at 45 degrees and fall to negligible corrections at backward angles, whereas the VV(CC, CL and LL)  contributions are negligible for all angles considered. 

The fact that the underlying transverse vector and axial-vector matrix elements are comparable in magnitude for the kinematics being explored in this work (hence the antineutrino suppression discussed above) has consequences for the uncertainties expected in the predictions made here for neutrinos. We recall from our treatment above that effects from MEC and their associated correlations are not present for these predictions, since they were ignored in analyzing the electron scattering cross sections. Note, however, that the error incurred by this may be less here than for electron scattering: the transverse contributions to the neutrino reaction cross sections involve both polar- and axial-vector matrix elements with both one-body, impulsive contributions (included here) and two-body MEC/correlation contributions (not included). From past studies one knows that the latter occur primarily in the transverse channel, but not in the longitudinal channel for electron scattering at the kinematics of interest here. That is, the vector MEC/correlation effects occur primarily in the transverse channel for QE and $\Delta$ kinematics at high energies. In contrast, due to the factor $\gamma_5$ that enters in axial-vector currents, the converse is true for axial-vector contributions --- accordingly, to leading order one does not expect large corrections of this type for the axial-vector contributions. As a consequence, the residual effect seen in Fig.~\ref{fdelta} at large negative $\psi'_\Delta$ and attributed to MEC and their associated correlations which amounted to roughly 10--15\% of the total cross section in that region of inelasticity measures the uncertainty in the {\em vector} contributions. Since these are only roughly 1/2 of the total for neutrinos, with the axial-vector transverse contributions accounting for the other half, the overall impact of the neglected MEC/correlation contributions is likewise only half as large, namely, providing less than 10\% uncertainty to the neutrino cross section predictions made here.

\begin{figure}[hbt]
\begin{center}
\hspace*{-1cm}\includegraphics[scale=0.5,clip]{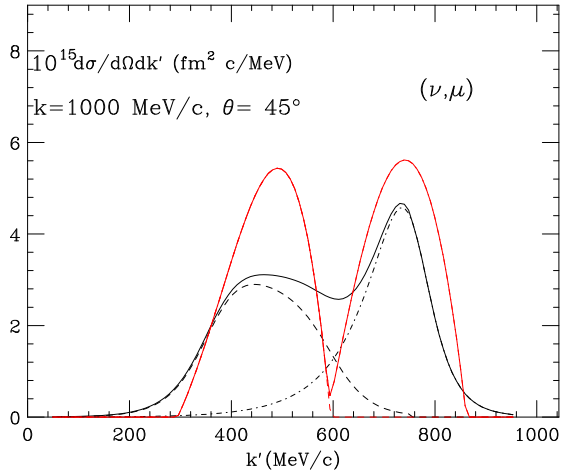}
\parbox{13cm}{\caption[]{Neutrino reaction cross sections as in 
Fig.~\ref{neutoncarbon} for $\theta=$ 45 degrees, showing a comparison of 
the full results obtained using the empirical scaling functions $f^{QE}$ and 
$f^\Delta$ obtained as discussed in the text with results obtained using the
RFG scaling function $f_{RFG}$ (heavier lines). The former lie somewhat 
lower and extend over a wider range in $k'$ than the latter.
}\label{rfgcomp}} 
\end{center} 
\end{figure}

For completeness in Fig.~\ref{rfgcomp} we show a comparison of a typical neutrino reaction cross section obtained using the full results with $f^{QE}$ and $f^\Delta$ deduced directly from electron scattering data with the cross section obtained using the RFG. As noted earlier, modeling done on the basis of mean-field theory is very similar to the RFG result, since at the relatively high energies of interest the dynamical effects embodied in an effective mass are expected to be small (most recent relativistic mean-field theory studies predict that $m^*$ reverts almost to the nucleon mass for the kinematics of interest). Likewise, relativized shell model predictions are close to the RFG predictions, and, moreover, RPA correlations are expected to be relatively small for the high energies involved. Thus, the RFG predictions effectively represent a larger set of models. As can be seen in the figure, all therefore differ significantly from the scaling predictions. 
Given the success of the scaling approach in studies of inclusive electron scattering 
for the kinematic region under study we expect that neutrino reaction 
cross sections also obtained using scaling ideas to be more robust 
than those based directly on existing models.

\section{Conclusions}
\label{sec:concl}

We begin by summarizing the approach followed in the present study. 

The first step has been to explore the scaling behavior of inclusive electron scattering for relatively high energies (several hundred MeV to a few GeV) in the kinematic region extending from the scaling region which lies below the QE peak, through the QE peak region and up to the peak where $\Delta$-excitation is the dominant process. Upon examining the longitudinal contribution one finds superscaling, {\it i.e.}, it is possible to find a scaling function $f^{QE}$ which, when plotted versus an appropriate scaling variable $\psi'_{QE}$, is seen to superscale. This means that the results are found to be independent both of the momentum transfer (scaling of the first kind) and of the particular nuclear species (scaling of the second kind). The assumption is made that this universal scaling function embodies the basic nuclear dynamics in the problem. Implicit in this, and borne out by modeling, is the observation that apparently MEC effects with their associated correlations and inelastic processes such as $\Delta$ excitation are not large contributions for the longitudinal response.

The second step has been to use the universal scaling function $f^{QE}$ to obtain that part of the transverse response that is due to impulsive processes, namely, those which arise only from elastic $eN$ scattering from nucleons in the nucleus. Upon subtracting this contribution from the total transverse response one finds a residual which is assumed to be due to the MEC/correlation effects and to inelastic $eN$ scattering from nucleons in the nucleus. From modeling we know that such effects are predominantly transverse and hence naturally occur in this channel, but are not very important in the longitudinal channel. Moreover, again from modeling of the various processes, we expect that the MEC/correlation effects are relatively small corrections for the kinematics of interest and accordingly attribute most of the residual to impulsive, inelastic $eN$ scattering, especially to contributions which arise from $N\to\Delta$ transitions. When the residual is analyzed in terms of an appropriate scaling function $f^\Delta$ by dividing the residual response by the elementary $N\to\Delta$ cross section and plotted versus an appropriate scaling variable $\psi'_\Delta$ which incorporates the kinematics of the inelastic transition we again find a reasonably successful new kind of scaling, at least for kinematics where the concept is expected to work. Specifically, this success is only found for excitations up to the peak of the region where the $\Delta$ dominates, but not beyond: this is not unexpected, since the approach taken in the present work has been tailored only to work when the $\Delta$ provides the basic driving process. Also, deviations are seen in the region where the QE and $\Delta$ responses overlap and there one does expect corrections from MEC and their associated correlations, while small, nevertheless to be necessary for a fully successful representation of the total response. A check is made by reassembling the complete inclusive $(e,e')$ cross section for the kinematic region of interest, typically finding errors of 10\% or less.

Thus, the first goal of this work has been met: we have found a very good representation of inclusive electron scattering at relatively high energies for the region of excitation extending up to the peak of the $\Delta$. It should be noted that direct modeling ({\it i.e.}, relativistic modeling, since non-relativistic approaches are known to fail badly for the kinematics of interest) yields electron scattering cross sections that are valid only at about the 25\% level or worse, while here our goal has been to use the scaling approach to do better.

The second major objective in this study has been to predict charge-changing neutrino and antineutrino cross sections for the same range of kinematics. Using the scaling approach in reverse, we take the empirically determined scaling functions $f^{QE}$ and $f^\Delta$ together with the appropriate $N\to N$ and $N\to\Delta$ charge-changing weak interaction cross sections to obtain the inclusive $\nu A$ and ${\bar\nu} A$ cross sections for the case of $^{12}$C. Given the above statements of where the approach taken in the present work should be valid we believe that these predictions should be the best currently available for few GeV neutrino reactions in the kinematic region that includes the full QE response and the $\Delta$ response up to its maximum. From our analyses of MEC contributions and 
how they enter in the relevant vector and axial-vector responses, we expect that 
corrections from such processes account for only about 10\% of the total cross section.
Note that, while our focus 
has been on the case of carbon, we know from our previous studies of scaling of the second kind that it is straightforward to produce predictions for other nuclei as well. In particular, while only selected predictions are given in the present work, further results at different kinematics and for other nuclei may be obtained by contacting our 
collaboration~\cite{Ingoemail}.

Finally, we mention our intentions for future work. Most straightforward will be to extend the scaling approach to obtain predictions for neutral current neutrino and antineutrino scattering from nuclei for similar kinematics; studies of this type are currently in progress. Also, given recent work on inelastic $eN$ processes and their role in $eA$ inclusive cross sections undertaken by the same collaboration, our intent is to explore a more microscopic model for all of these reactions. While this project is relatively straightforward as well, what is lacking before it can be realized is a completion of our on-going studies of MEC effects, together with the correlations they require due to gauge invariance; studies of this type are also in progress. Finally, there is the issue of explaining the specific nature of the scaling functions themselves. While there are indications that contributions from high Fourier components in the nuclear ground state are probably responsible for the detailed nature of these functions, a fully satisfactory relativistic treatment of them is presently lacking. Until one becomes available it appears that the best approach is to take the scaling functions directly from experiment, as we have done in the present work.

%***************************
\section*{Acknowledgments}
%***************************

This work was partially supported by funds provided by DGI (Spain) and
FEDER funds, under Contracts Nos. BFM2002-03218, BFM2002-03315 and
FPA2002-04181-C04-04 and by the Junta de Andaluc\'{\i}a,
and by the INFN-CICYT collaboration agreement (project 
``Study of relativistic dynamics in electron and neutrino scattering'').
It was also supported in part (TWD) by the U.S. Department of Energy
under cooperative research agreement No. DE-FC02-94ER40818 and in part (IS)
by the Schweizerische Nationalfonds. The authors wish to thank members of the various 
neutrino oscillation collaborations for very helpful discussions of on-going 
neutrino oscillation experiments.

\newpage

\section*{Appendix}
\label{sec:appen}

Performing the traces in Eq.~(\ref{Tr}) one gets:
\ba
w_{1V} &=&
\! \left(\frac{{\cal T}^2}{3}\right) 
\frac{1}{4\mu_\Delta^2} 
\ [4\ \tau + (\mu_\Delta-1)^2]
\Bigg\{
\Bigg[16\ \tau^2 +8\tau(\mu_\Delta+1) + (\mu_\Delta+1)^2(3\mu_\Delta^2+1)
\Bigg]
C_3^{V2}
\nonumber\\
&+&
\mu_\Delta^2 
\Bigg[ 
(1 +4\ \tau - \mu_\Delta^2)C_4^V+(1 - 4\ \tau - \mu_\Delta^2)C_5^V
\Bigg]^2
\nonumber\\
&+&
\mu_\Delta
C_3^V (1+4\tau-2\mu_\Delta-3\mu_\Delta^2)
\Bigg[ 
(1 + 4\ \tau - \mu_\Delta^2)C_4^V+(1 - 4\ \tau - \mu_\Delta^2)C_5^V
\Bigg]
\Bigg\}
\nonumber\\
\label{24}
\ea
\ba
w_{2V} &=&
\! \left(\frac{{\cal T}^2}{3}\right) \frac{4\tau}{\mu_\Delta^2} 
\Bigg\{
(1 + 4\ \tau+ 3\ \mu_\Delta^2) 
C_3^{V2}
\nonumber\\
&+&
\Bigg[4\ \tau + (\mu_\Delta-1)^2\Bigg]
\Bigg[\mu_\Delta^2(C_4^V+C_5^V)^2+4\tau C_5^{V2}\Bigg]
\nonumber\\
&+&
\mu_\Delta \Bigg[ (1 + 4\ \tau - 4\mu_\Delta+3\mu_\Delta^2)(C_4^V+C_5^V)
+8\ \tau\ C_5^V\Bigg]
C_3^V 
\Bigg\}
\label{25}
\ea
for the vector contributions, whereas they yield 
\ba
w_{1A} &=&
\! \left(\frac{{\cal T}^2}{3}\right) \frac{1}{4\mu_\Delta^2} 
\ [4\ \tau + (\mu_\Delta+1)^2]
\Bigg\{
\Bigg[16\ \tau^2 -8\tau(\mu_\Delta-1) + (\mu_\Delta-1)^2(3\mu_\Delta^2+1)
\Bigg]
C_3^{A2}
\nonumber\\
&+&
\mu_\Delta^2 
\Bigg[ 
(1 +4\ \tau - \mu_\Delta^2)C_4^A-2\ C_5^V
\Bigg]^2
\nonumber\\
&+&
\mu_\Delta
C_3^A (1+4\tau+2\mu_\Delta-3\mu_\Delta^2)
\Bigg[ 
(1 + 4\ \tau - \mu_\Delta^2)C_4^A-2\ C_5^A
\Bigg]
\Bigg\}
\label{26}
\ea
\ba
w_{2A} &=&
\! \left(\frac{{\cal T}^2}{3}\right) \frac{4\tau}{\mu_\Delta^2} 
\Bigg\{
(1 + 4\ \tau+ 3\ \mu_\Delta^2) 
C_3^{A2}
\nonumber\\
&+&
\Bigg[4\ \tau + (\mu_\Delta+1)^2\Bigg]
\Bigg(\mu_\Delta^2\ C_4^{A2}+\frac{C_5^{A2}}{4\tau}\Bigg)
\nonumber\\
&+&
\mu_\Delta\Bigg[(1+4\tau+4\mu_\Delta+3\mu_\Delta^2)C_4^A+2\ C_5^A\Bigg]
C_3^A 
\Bigg\}
\label{27}
\ea
\ba
u_{1A} &=&
\! -\left(\frac{{\cal T}^2}{3}\right) \frac{1}{16\ \tau\ \mu_\Delta^2}
\ [4\ \tau + (\mu_\Delta+1)^2]
(C_5^A - 4\ \tau\ C_6^A)
\nonumber
\\
&\times&
\Bigg\{
\ C_5^A\ \Bigg[48\ \tau^2 - (\mu_\Delta^2-1)^2 + 8\ \tau\ (\mu_\Delta^2+1)\Bigg]
\nonumber
\\
&+& 
            4\ \tau\ \Bigg[4\ \mu_\Delta\ (C_3^A + \mu_\Delta\ C_4^A)
                        \ (4\ \tau + \mu_\Delta^2-1) 
\nonumber
\\
&-&
                  C_6^A\ [16\ \tau^2 + (\mu_\Delta^2-1)^2 + 
                        8\ \tau\ (\mu_\Delta^2+1)]\Bigg]\Bigg\}  
\label{28}
\ea
\ba
u_{2A} &=&
\! \left(\frac{{\cal T}^2}{3}\right) \frac{1}{\ 4 \tau\mu_\Delta^2}\ [4\ \tau + (\mu_\Delta+1)^2]
(C_5^A - 4\ \tau\ C_6^A\ )
\nonumber
\\
&\times&
\ [8\ \tau\ \
\mu_\Delta\ (C_3^A+\ \mu_\Delta\ C_4^A) + 
              C_5^A\ (4\ \tau - 
                    \mu_\Delta^2+1)]
\label{29}
\ea
for the axial-vector contributions and

\ba
w_{3} &=&
\! \left(\frac{{\cal T}^2}{3}\right) 
\frac{1}{2\ \mu_\Delta^2}
\Bigg\{\mu_\Delta\Bigg[2\ C_5^A
-(4\tau-\mu_\Delta^2+1)\ C_4^A \Bigg]
\nonumber\\
&\times&
\Bigg[ 
(1 + 4\ \tau + 4\mu_\Delta+3 \mu_\Delta^2)\ C_3^V
-\mu_\Delta
[(4\tau-\mu_\Delta^2+1)\ C_4^V
-(4\tau+\mu_\Delta^2-1)\ C_5^V ]
\Bigg]
\nonumber\\
&-& 
C_3^A \ \Bigg[2(1+8\tau+16\tau^2+2\mu_\Delta^2-3\mu_\Delta^4)\ C_3^V
\nonumber\\
&+&\mu_\Delta(1+4\tau-4\mu_\Delta+3\mu_\Delta^2)
[(4\tau-\mu_\Delta^2+1)\ C_4^V
-(4\tau+\mu_\Delta^2-1)\ C_5^V ]
\Bigg]
\Bigg\}
\label{30}
\ea
for the interference pieces. 
Since the axial-vector current is not conserved, $w_1$ and $w_2$ are not sufficient
to set up the $AA$ hadronic tensor: hence two extra functions $u_{1A}$ and
$u_{2A}$, which vanish if $C_5^A=4\tau C_6^A$, come out from the traces.

For the empirical functions entering above we take~\cite{Alvarez-Ruso:1998hi}
\ba
C_3^V(\tau)&=&\frac{2.05}{(1+|Q^2|/0.54 GeV^2)^2}
\\
C_4^V(\tau)&=&-\frac{C_3^V}{\mu_\Delta}
\\
C_5^V(\tau)&=&0
\\
C_3^A(\tau)&=&0
\\
C_4^A(\tau)&=&-0.3\,\left(1-\frac{1.21 |Q^2|}{2 GeV^2+|Q^2|}\right)
\left(1+\frac{|Q^2|}{(1.28)^2 GeV^2}\right)^{-2}
\\
C_5^A(\tau)&=&1.2\,\left(1-\frac{1.21 |Q^2|}{2 GeV^2+|Q^2|}\right)
\left(1+\frac{|Q^2|}{(1.28)^2 GeV^2}\right)^{-2}
\\
C_6^A(\tau)&=&C_5^A(\tau)\frac{m_N^2}{m_\pi^2+|Q^2|}
=\frac{C_5^A}{4\tau+\mu_\pi^2}~,
\ea
which, inserted into Eqs.~(\ref{25}-\ref{30}), lead to
\ba
 w_{1V} &=&
\! \left(\frac{{\cal T}^2}{3}\right) \frac{1}{4\mu_\Delta^2} 
\ [4 \tau + (\mu_\Delta+1)^2]^2
\ [4 \tau + (\mu_\Delta-1)^2]\ 
C_3^{V2}
\label{eq:w1v}
\\
w_{2V} &=&
\! \left(\frac{{\cal T}^2}{3}\right) \frac{4\tau}{\mu_\Delta^2} 
\ [4\ \tau+ (\mu_\Delta+1)^2] 
C_3^{V2}
\label{eq:w2v}
\ea
for the vector contributions
\ba
w_{1A} &=&
\! \left(\frac{{\cal T}^2}{3}\right) \frac{1}{4} 
\ [4 \tau + ( \mu_\Delta+1)^2]
\ [(4\tau-\mu_\Delta^2+1)C_4^A-2 C_5^A]^2
\label{eq:w1a}
\\
w_{2A} &=&
\! \left(\frac{{\cal T}^2}{3}\right) \frac{1}{\mu_\Delta^2} 
\ [4\ \tau + \ (\mu_\Delta+1)^2]
\ (4\tau\mu_\Delta^2 C_4^{A2}+C_5^{A2})
\label{eq:w2a}
\\
u_{1A} &=&
\! -\left(\frac{{\cal T}^2}{3}\right) 
\frac{1}{16\tau\mu_\Delta^2}\ [4\ \tau + (\mu_\Delta+1)^2]
(C_5^A - 
            4\tau C_6^A\ )
\nonumber
\\
&\times&
\ \Bigg\{ \ \Bigg[48
\tau^2 - (\mu_\Delta^2-1)^2 + 8\tau (\mu_\Delta^2+1)\Bigg] C_5^A
\nonumber
\\
&+& 
            4 \tau\ \Bigg[4 \mu_\Delta^2  \  
                        (4 \tau + \mu_\Delta^2-1) C_4^A 
%\nonumber
%\\
%&-&
-
                  \ [16 \tau^2 + (\mu_\Delta^2-1)^2 + 
                        8 \tau\ (\mu_\Delta^2+1)] C_6^A \Bigg]\Bigg\}  
%\nonumber\\
\label{eq:u1a}
\\
 u_{2A} &=&
\! \left(\frac{{\cal T}^2}{3}\right) 
\frac{1}{\ 4 \tau\mu_\Delta^2}\ [4 \tau + (\mu_\Delta+1)^2]
\ (C_5^A - 4 \tau\ C_6^A)
%\nonumber
%\\
%\nonumber
%\\
%&\times&
\ [8 \tau 
\mu_\Delta^2 C_4^A + 
              \ (4\ \tau - 
                    \mu_\Delta^2+1) C_5^A]
\nonumber\\
\label{eq:u2a}
\ea
for the axial-vector contributions, and
\ba
w_{3} &=&
\! \left(\frac{{\cal T}^2}{3}\right) \frac{C_3^V}{\mu_\Delta} 
\ [4 \tau + ( \mu_\Delta+1)^2]
\ [2 C_5^A-(4\tau-\mu_\Delta^2+1) C_4^A] 
%\nonumber\\
\label{eq:w3va}
\ea
for the V/A interference.

One finds that $u_{1A}$ and $u_{2A}$ (which arise from
PCAC) are negligible, whereas the other functions are all significant. The latter are seen to fall strongly with increasing $q$.

\end{document}